\pdfoutput=1

\documentclass[11pt,twoside,a4paper,cmspaper,final,collab]{cms-tdr}
\def\svnVersion{473694P}\def\cmsCernNoTag{CERN-EP-2018-124}\def\cmsCernDate{\today}\def\cmsMessage{Published in the Journal of High Energy Physics as \href{http://dx.doi.org/10.1007/JHEP08(2018)113}{\doi{10.1007/JHEP08(2018)113.}}}

\begin{document}\cmsNoteHeader{HIG-16-018}
\providecommand{\hltdijmed}  {HLT\_Jet60Eta1p7\_Jet53Eta1p7\_DiBTagIP3DFastPV}
\providecommand{\hltdijhigh} {HLT\_Jet80Eta1p7\_Jet70Eta1p7\_DiBTagIP3DFastPV}
\providecommand{\hltdijvhi}  {HLT\_Jet160Eta2p4\_Jet120Eta2p4\_DiBTagIP3DFastPVLoose}

\providecommand{\hltdijlowa}{HLT\_CentralJet46\_BTagIP3D\_CentralJet38\_BTagIP3D}
\providecommand{\hltdijlow}{HLT\_CentralJet46\_CentralJet38\_DiBTagIP3D}
\providecommand{\hltdijhi}{HLT\_CentralJet60\_CentralJet53\_DiBTagIP3D}
\providecommand{\hlttrij}{HLT\_CentralJet46\_CentralJet38\_CentralJet20\_DiBTagIP3D}

\providecommand{\ldijmed}  {L1\_DoubleJetC44\_Eta1p74\_WdEta4}
\providecommand{\ldijhigh} {L1\_DoubleJetC56\_Eta1p74\_WdEta4}
\providecommand{\ldijvhi}  {L1\_SingleJet128}

\providecommand{\ldijlow}{L1\_DoubleJet36\_Central}
\providecommand{\ldijhi} {L1\_DoubleJet44\_Central}
\providecommand{\ltrij}  {L1\_TripleJet\_36\_36\_12\_Central}

\providecommand{\bb}{$b\bar{b}$}

\providecommand{\pt}{$p_T$}
\providecommand{\ptj}{$p_{T,jet}$}
\providecommand{\etaj}{$\eta_{jet}$}
\providecommand{\detajj}{$\Delta\eta_{jet,jet\prime}$}
\providecommand{\clsnf}{$95\%$ CL}

\hyphenation{had-ron-i-za-tion}
\hyphenation{cal-or-i-me-ter}
\hyphenation{de-vices}
\RCS$HeadURL: svn+ssh://svn.cern.ch/reps/tdr2/papers/HIG-16-018/trunk/HIG-16-018.tex $
\RCS$Id: HIG-16-018.tex 473025 2018-08-27 11:46:22Z chayanit $

\newlength\cmsTabSkip\setlength{\cmsTabSkip}{1ex}
\newcommand{\mA}{\ensuremath{m_{\PSA}}\xspace}
\newcommand{\mh}{\ensuremath{m_{\Ph}}\xspace}
\newcommand{\mH}{\ensuremath{m_{\PH}}\xspace}
\newcommand{\mAH}{\ensuremath{m_{\PSA/\PH}}\xspace}
\newcommand{\mPhi}{\ensuremath{m_{\phi}}\xspace}
\newcommand{\mjj}{\ensuremath{M_{12}}\xspace}
\newcommand{\mhmax}{\ensuremath{m_{\Ph}^{\text{max}}}\xspace}
\newcommand{\mhmodp}{\ensuremath{m_{\Ph}^{\text{mod+}}}\xspace}
\newcommand{\mhmodm}{\ensuremath{m_{\Ph}^{\text{mod--}}}\xspace}
\newcommand{\mZ}{\ensuremath{m_{\PZ}}\xspace}
\ifthenelse{\boolean{cms@external}}{\providecommand{\NA}{\ensuremath{\cdots}\xspace}}{\providecommand{\NA}{\ensuremath{\text{---}}\xspace}}
\newcommand{\cosba}{\ensuremath{\cos(\beta-\alpha)}\xspace}
\ifthenelse{\boolean{cms@external}}{\providecommand{\CL}{C.L.\xspace}}{\providecommand{\CL}{CL\xspace}}
\newcommand{\CP}{\ensuremath{CP}\xspace}

\newlength\cmsFigWidth
\ifthenelse{\boolean{cms@external}}{\setlength\cmsFigWidth{0.85\columnwidth}}{\setlength\cmsFigWidth{0.4\textwidth}}
\ifthenelse{\boolean{cms@external}}{\providecommand{\cmsLeft}{top\xspace}}{\providecommand{\cmsLeft}{left\xspace}}
\ifthenelse{\boolean{cms@external}}{\providecommand{\cmsRight}{bottom\xspace}}{\providecommand{\cmsRight}{right\xspace}}

\cmsNoteHeader{HIG-16-018}

\title{Search for beyond the standard model Higgs bosons decaying into a \bbbar\ pair in \Pp\Pp\ collisions at $\sqrt{s} = 13\TeV$}

\date{\today}

\abstract{A search for Higgs bosons that decay into a bottom quark-antiquark pair and are accompanied by at least one additional bottom quark is performed with the CMS detector. The data analyzed were recorded in proton-proton collisions at a centre-of-mass energy of $\sqrt{s} = 13\TeV$ at the LHC, corresponding to an integrated luminosity of 35.7\fbinv. The final state considered in this analysis is particularly sensitive to signatures of a Higgs sector beyond the standard model, as predicted in the generic class of two Higgs doublet models (2HDMs). No signal above the standard model background expectation is observed. Stringent upper limits on the cross section times branching fraction are set for Higgs bosons with masses up to 1300\GeV. The results are interpreted within several MSSM and 2HDM scenarios.}

\hypersetup{
pdfauthor={CMS Collaboration},
pdftitle={Search for beyond the standard model Higgs bosons decaying into a bbbar pair in pp collisions at sqrt(s) = 13 TeV},
pdfsubject={CMS},
pdfkeywords={CMS, physics, Higgs, MSSM}}

\maketitle

\section{Introduction}

In the standard model (SM), a Higgs boson at a mass of 125\GeV has a large coupling to \cPqb{} quarks via Yukawa interactions. Its production in association with \cPqb{} quarks and subsequent decay into \cPqb{} quarks at the CERN LHC is, however, difficult to detect because of the high rate of heavy-flavour multijet production. There are, nevertheless, models beyond the SM that predict an enhancement of Higgs boson production in association with \cPqb{} quarks, which motivate the search for such processes.

Prominent examples of models beyond the SM are the two Higgs doublet model (2HDM)~\cite{Branco:2011iw}, which contains two scalar Higgs doublets, as well as one particular realization within the minimal supersymmetric extension of the SM (MSSM)~\cite{Nilles19841}. These result in two charged Higgs bosons, \PHpm\, and three neutral ones, jointly denoted as $\phi$.
Among the latter are, under the assumption that \CP\ is conserved, one \CP-odd (A), and two \CP-even (h, \PH) states, where \Ph usually denotes the lighter \CP-even state. For the purpose of this analysis, the boson discovered in 2012 with a mass near 125\GeV~\cite{Aad:2012gk,Chatrchyan:2012gu,Chatrchyan:2013lba,Sirunyan:2017exp} is interpreted as \Ph, whose mass is thus constrained to the measured value. The two heavier neutral states, \PH and A, are the subject of the search presented here.

In the 2HDM,  flavour changing neutral currents at tree level can be suppressed by introducing discrete symmetries, which restrict the choice of Higgs doublets to which the fermions can couple. This leads to four types of models with natural flavour conservation at tree level:
\begin{itemize}
	\item {\bf type-I}: all charged fermions couple to the same doublet;
\item {\bf type-II}: up-type quarks (\cPqu, \cPqc, \cPqt) couple to one doublet, down-type
	fermions (\cPqd, \cPqs, \cPqb, \Pe, \Pgm, \Pgt) couple to the other. This
	structure is also implemented in the MSSM;
	\item {\bf lepton-specific}: all charged leptons couple to one doublet, all quarks couple
	to the other;
	\item {\bf flipped}:  charged leptons and up-type quarks couple to one doublet, down-type
	quarks to the other.
\end{itemize}
While until
now the type-I and -II models have been most intensively tested, the flipped model is remarkably unexplored
from the experimental side. The $\PSA/\PH
\to \bbbar{}$ decay mode is ideally suited to constrain this
model due to the large branching fraction of the Higgs boson into \cPqb{} quarks.

The \CP-conserving 2HDMs have seven free parameters. They can be chosen as the Higgs boson masses (\mh, \mH, \mA, $m_{\PHpm}$),
the mixing angle between the \CP-even Higgs bosons ($\alpha$), the ratio of the vacuum expectation values of the two doublets ($\tanb = v_2/v_1$),
and the parameter that potentially mixes the two Higgs doublets ($m_{12}$). For $\cosba \to 0$, the light \CP-even Higgs boson (h) obtains properties indistinguishable
from the SM Higgs boson with the same mass in all four types of models listed above~\cite{Branco:2011iw}.

The MSSM Higgs sector has the structure of a type-II 2HDM.
The additional constraints given by the fermion-boson symmetry fix all mass relations between the Higgs bosons and the angle $\alpha$
at tree level, reducing the number of parameters at this level to only two.
These parameters are commonly chosen as the mass of the pseudoscalar Higgs boson, \mA, and \tanb.
After the Higgs boson discovery at the LHC, MSSM benchmark scenarios have been refined to match
the experimental data and to reveal characteristic features of certain regions of the parameter space~\cite{Carena:2013qia,Carena:2005ek}.
Considered in this analysis are the $\mhmodp$, the hMSSM~\cite{Djouadi:2013uqa}, the light stau ($\sTau$), and the light stop ($\sTop$) scenarios~\cite{Carena:2013qia}.

The $\mhmodp$ scenario is a modification of the $\mhmax$ scenario, which was originally defined to give conservative exclusion bounds on $\tanb$ in the LEP Higgs boson searches~\cite{Schael:2006cr,Carena:2002qg,Heinemeyer:1999zf}. It has been modified such that the mass of the lightest \CP-even state, \mh, is compatible with the mass of the observed boson within $\pm3\GeV$~\cite{Degrassi:2002fi,Allanach:2004rh} in a large fraction of the considered parameter space~\cite{Carena:2013qia}. The hMSSM approach~\cite{Maiani:2013hud,Djouadi:2013uqa,Djouadi:2015jea} describes the MSSM Higgs sector in terms of just \mA and \tanb, given the experimental knowledge of \mZ and \mh. It defines a largely model-independent scenario, because the predictions for the properties of the MSSM Higgs bosons do
not depend on the details of the supersymmetric sector~\cite{Bagnaschi:2039911}.
Further variations of the supersymmetric sector are implemented in the light $\sTau$ and light $\sTop$ scenarios~\cite{Carena:2013qia}, which are also designed such that the light scalar \Ph is compatible with the measured Higgs boson mass~\cite{Aad:2015zhl}.

For \tanb values larger than one, the
couplings of the Higgs fields to \cPqb{} quarks are enhanced both in the flipped and the type-II models, and thus also in the MSSM.
Furthermore, there is an approximate mass degeneracy between the \PSA and \PH bosons in the MSSM for the studied range of \mA. For the 2HDM scenarios considered in this paper, such a degeneracy will be imposed. These effects enhance the
combined cross section for producing these Higgs bosons in association with \cPqb{}
quarks by a factor of up to 2 \tanbsq with respect to the SM. The decay $\PSA/\PH\to\bbbar$
is expected to have a high branching fraction, even at large values of the Higgs boson mass and $\abs{\cosba}$~\cite{Heinemeyer:2013tqa}.

The most stringent constraints on the MSSM parameter \tanb so far, with exclusion limits in the range 4--60 in the mass interval of 90--1600\GeV, have been obtained in measurements at the LHC in the $\phi\to\tau\tau$ decay
mode~\cite{ref:Chatrchyan201268,Aad:2012cfr,Khachatryan:2014wca,Aad:2014vgg,Aaboud:2017sjh,Sirunyan:2018zut}. Preceding limits have been obtained by the LEP~\cite{Schael:2006cr} and the Tevatron experiments~\cite{Aaltonen:2009vf,Abazov:2008hu,Abazov2012569}.
The $\phi\to\Pgm\Pgm$ decay mode has been investigated as well~\cite{Aad:2012cfr,CMS:2015ooa,Sirunyan:2017uvf}.

In the $\phi\to\bbbar$ decay mode, searches have initially been
performed at LEP~\cite{Schael:2006cr} and by the CDF and \DZERO Collaborations~\cite{PhysRevD.86.091101} at
the Tevatron collider. At the LHC, the only analyses
in this channel with associated \cPqb{} jets have also been performed by the CMS Collaboration using
the 7 and 8\TeV data~\cite{Chatrchyan:2013qga,Khachatryan:2015tra}.
In the absence of any signal, limits on the
$\Pp\Pp \to \cPqb\phi(\to \bbbar) + \mathrm{X}$ cross section
have been derived in the 90--900\GeV mass range. The
combined 7 and 8\TeV data analyses
translate into upper bounds on \tanb between 14 and 50 in the
Higgs boson mass range of 100--500\GeV, assuming the $\mhmodp$ scenario of
the MSSM.

The ATLAS and CMS Collaborations have performed extensive 2HDM interpretations of measurements in different production and decay channels, in particular also in the $\PSA \to \PZ\Ph$, $\Ph \to \bbbar$ decay mode~\cite{Aad:2015wra,Aaboud:2017cxo,Khachatryan:2015lba}. The ATLAS interpretation~\cite{Aaboud:2017cxo} also covers the flipped scenario, and the 2HDM interpretations reported in this paper are compared to these.

\begin{figure}[htbp]
	\centering
	\includegraphics[width=0.28\textwidth]{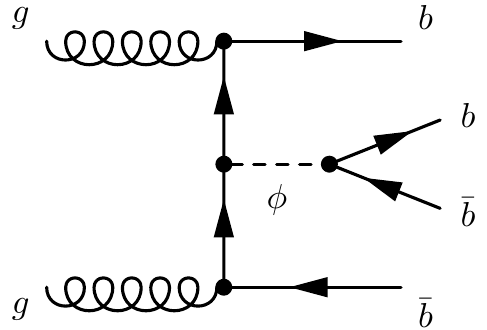}\hfill
	\includegraphics[width=0.28\textwidth]{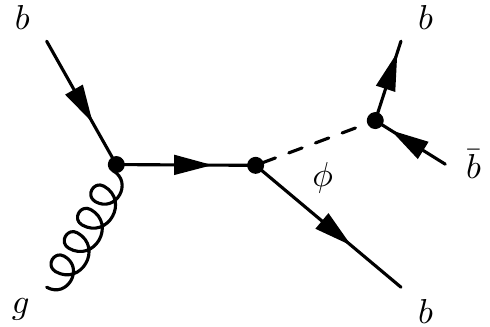}\hfill
	\includegraphics[width=0.28\textwidth]{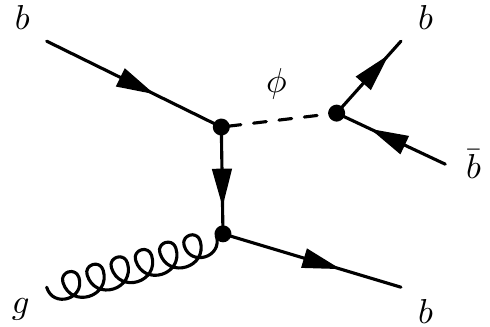}
	
	\caption{Example Feynman diagrams for the signal processes.
	}
	\label{fig:signalprocesses}
\end{figure}

With the proton-proton (\Pp\Pp) collision data set corresponding to an integrated luminosity of 35.7\fbinv collected at a centre-of-mass energy of $\sqrt{s} = 13\TeV$ in 2016, the sensitivity to key model parameters with respect to previous CMS searches is significantly extended. The analysis focuses on neutral Higgs bosons \PSA and \PH with masses $\mAH \geq 300\GeV$ that are produced in association with at least one \cPqb{} quark and decay to $\bbbar$, as shown by the diagrams in Fig.~\ref{fig:signalprocesses}. The signal signature therefore comprises final states characterized by at least three \cPqb\ quark jets ("\cPqb\ jets"), and the dominant background is multijet production. A fourth \cPqb\ jet is not explicitly required, since due to the process topology the majority of the signal events are found to have at most three \cPqb\ jets within the acceptance of this analysis. Events are selected by dedicated triggers that identify \cPqb\ jets already during data taking. This helps significantly to suppress the large rate of multijet production, while maintaining sensitivity to the signal process. The analysis searches for a peak in the invariant mass distribution, $\mjj$, of the two \cPqb\ jets with the highest transverse momentum \pt values, which originate from the Higgs boson decay in about 66\% of all cases at $\mAH = 300\GeV$, increasing up to 75\% for $\mAH \geq 700\GeV$. The dominant background is the production of heavy-flavour multijet events containing either three \cPqb\ jets, or two \cPqb\ jets plus a third jet originating from either a charm quark, a light-flavour quark, or a gluon, which is misidentified as a \cPqb\ jet.

\section{The CMS detector}

The central feature of the CMS apparatus is a superconducting solenoid
of 6\unit{m} internal diameter, providing a magnetic field of
3.8\unit{T}.  Within the field volume, the inner tracker is formed by
a silicon pixel and strip tracker.  It measures charged particles
within the pseudorapidity range $\abs{\eta}< 2.5$.
The tracker
provides a transverse impact parameter resolution of approximately 15\mum and
a resolution on \pt  of about 1.5\% for particles with $\pt = 100\GeV$.
Also inside the field volume are a crystal electromagnetic
calorimeter, and a brass and scintillator hadron calorimeter.
Forward calorimetry extends the coverage provided by the barrel
and endcap detectors up to $\abs{\eta}<5$. Muons are
measured in gas-ionization detectors embedded in the steel flux-return
yoke, in the range $\abs{\eta}< 2.4$, with detector
planes made using three technologies: drift tubes, cathode strip
chambers, and resistive-plate chambers. Matching muons to tracks
measured in the silicon tracker results in a \pt
resolution between 1 and 10\%, for \pt values up to 1\TeV.
A detailed description of the CMS detector,
together with a definition of the coordinate system used and the relevant
kinematic variables,
can be found in Ref.~\cite{Chatrchyan:2008zzk}.

\section{Event reconstruction and simulation}

A particle-flow algorithm~\cite{Sirunyan:2017ulk} aims to
reconstruct and identify all particles in
the event, \ie electrons, muons, photons, and charged and
neutral hadrons, with an optimal combination of all CMS detector
systems.

The reconstructed vertex with the largest value of summed physics-object $\pt^2$ is taken to be the primary \Pp\Pp\ interaction vertex. The physics objects chosen are those that have been defined using information from the tracking detector, including jets, the associated missing transverse momentum, which is taken as the negative vector sum of the \pt of those jets, and charged leptons.

Jets are clustered from the reconstructed particle-flow candidates using the anti-\kt
algorithm~\cite{Cacciari:2008gp,Cacciari:2011ma} with a distance parameter of 0.4.
Each jet is required to pass
dedicated quality criteria to suppress the impact of instrumental
noise and misreconstruction. Contributions from additional
\Pp\Pp\ interactions within the same or neighbouring bunch crossing (pileup)
affect the jet momentum measurement. To mitigate this effect,
charged particles associated with other vertices than the reference primary
vertex are discarded before jet reconstruction~\cite{CMS-PAS-JME-14-001}, and residual
contributions (\eg from neutral particles)
are accounted for using a jet-area based
correction~\cite{Cacciari2008119}.
Subsequent jet energy corrections are derived from simulation, and are confirmed
with in situ measurements of the energy balance in dijet, multijet, and \Z/\GAMJET{}
events~\cite{Khachatryan:2016kdb}.

For the off\/line identification of \cPqb\ jets, the combined secondary
vertex (CSVv2) algorithm \cite{BTV-16-002}
is used. This algorithm combines information on track impact parameters and secondary
vertices within a jet into an artificial neural network classifier that provides
separation between \cPqb\ jets and jets of other flavours.

Simulated samples of signal and background events were produced using different event generators
and include pileup events. The MSSM Higgs boson signal samples, $\Pp\Pp\to\bbbar\phi$+X with
$\phi\to\bbbar$, were produced at leading order (LO) in the 4-flavour scheme with \PYTHIA 8.212~\cite{Sjostrand:2015gs}.
Comparing this prediction to computationally expensive next-to-leading order (NLO)
calculations~\cite{Dittmaier:2012vm} generated using \MGvATNLO
in version 2.3.0~\cite{Alwall:2014hca,Wiesemann:2014ioa}, we find a very good
agreement in the shapes of the leading dijet invariant mass distribution, $\mjj$,
while the selection efficiency is up to 10\% lower when using the NLO prediction.
We correct the NLO effect by applying mass-dependent correction factors to the LO signal samples
and assign a corresponding systematic uncertainty in the final results.
Multijet background events from quantum chromodynamics (QCD)
processes have been simulated with the \MGvATNLO event
generator~\cite{Frederix:2009yq,Hirschi:2011pa} using the 5-flavour scheme and MLM merging~\cite{Alwall:2007fs};
they are used for studying qualitative features but not for a quantitative background prediction.
The NNPDF 3.0~\cite{Ball:2011mu} parton distribution functions (PDFs) are used in all generated samples.
For all generators, fragmentation,
hadronization, and the underlying event have been modelled using \PYTHIA with tune CUETP8M1~\cite{Khachatryan:2015pea}.
The response of the CMS detector is modelled with the \GEANTfour toolkit~\cite{Agostinelli:2002hh}.

\section{Trigger and event selection}
\label{sec:trigger}

A major challenge to this search is posed by the huge hadronic interaction
rate at the LHC. This is addressed with a dedicated trigger scheme~\cite{Khachatryan:2016bia}, especially designed
to suppress the multijet background.
Only events with at least two jets in the
range of $\abs{\eta} \leq 2.4$ are selected.
The two leading jets
are required to have $\pt>100\GeV$, and an event is accepted only if the
absolute value of the difference in pseudorapidity, $\Delta \eta$, between any two jets fulfilling the \pt
and $\eta$ requirements, is less than or equal to 1.6.
The tight online requirements on the
opening angles between jets are introduced to reduce the trigger rates, while preserving high efficiency
in the probed mass range of the Higgs bosons.
At trigger level, \cPqb\ jets are identified using the CSVv2 algorithm with slightly tighter requirements
than for the off\/line analysis. At least two jets
in the event must satisfy the online \cPqb\ tagging criteria.

The efficiency of the jet \pt requirements in the trigger is derived from
data collected with prescaled single-jet triggers with lower threshold. The efficiency in data and simulation is measured
as a function of jet $\pt$ and $\eta$. The differences between the two are corrected for in the
analysis of the simulated samples. The online \cPqb\ tagging efficiencies relative to the off\/line \cPqb\ tagging
selection are obtained from data using prescaled dijet triggers with a single b-tag requirement.
A tag-and-probe method is employed to determine the efficiency as a function of $\pt$ and $\eta$ of the jets. Both leading jets are required to pass offline selection criteria including \cPqb\ tagging requirements similar to the final event selection described below. The second-leading \cPqb\ jet must always pass the online \cPqb\ tagging requirement to ensure that it has fired the trigger. The fraction of the first leading \cPqb\ jets that also satisfy the online \cPqb\ tagging requirements is a direct measure of the relative online \cPqb\ tagging efficiency. Relative efficiencies are found to range from above 80\% for $\pt \approx 100\GeV$ to around 50\% for $\pt \approx 900\GeV$, averaged over $\eta$.

The off\/line selection requires at least two jets with $\pt>100\GeV$ and another one with $\pt>40\GeV$, which all need to satisfy $\abs{\eta}\leq 2.2$.
The $\eta$ selection is applied to benefit from optimal \cPqb\ tagging performance.
The three leading jets have to pass the CSVv2 \cPqb\ tagging requirement of the medium working point~\cite{BTV-16-002}.
This working point features a 1\% probability for light-flavour jets (attributed to \cPqu, \cPqd, \cPqs, or \cPg\ partons) to be misidentified as \cPqb\ jets, and has a \cPqb\ jet identification efficiency of about 70\%.
The separation between the two leading jets in $\eta$ has to be less than 1.55, and
a minimal pairwise separation of $\Delta R>1$ between each two of the three leading jets is imposed to suppress
background from \bbbar{} pairs arising from gluon splitting. This sample is referred to
as ``triple \cPqb\ tag'' sample in the following.

\section{Signal modeling}
\label{sec:signaltempl}
A signal template for the \mjj\ distribution is obtained for each Higgs
boson mass considered by applying the full selection to the
corresponding simulated signal data set, for nominal masses in the
range of 300--1300\GeV.
The sensitivity of this analysis does not extend down to cross
sections as low as that of the SM Higgs boson. Thus,
a signal model with a single mass
peak is sufficient. This is in contrast to the $\phi\to\tau\tau$
analysis~\cite{Sirunyan:2018zut}, where the signal model comprises
the three neutral Higgs bosons of the MSSM, one of which is SM-like.

\begin{figure}[htpb]
	\centering
	\includegraphics[width=0.66\textwidth]{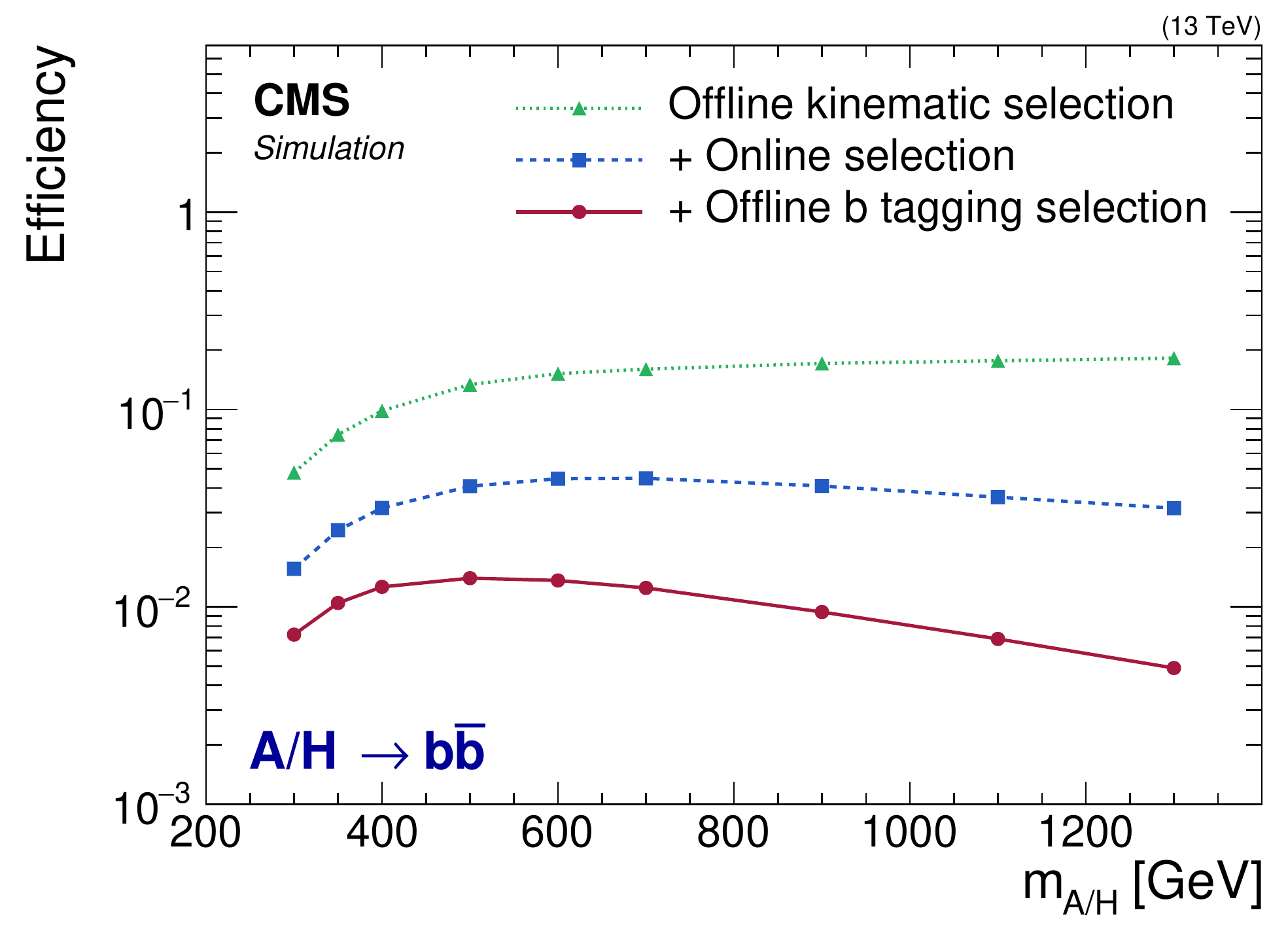}
	\caption{
		Signal efficiency as a function of the Higgs boson mass after different stages of event selection.
	}
	\label{fig:signalHEfficiency}
\end{figure}

The signal efficiency for each Higgs boson mass point is obtained from simulation and
shown in Fig.~\ref{fig:signalHEfficiency}.
A scale factor for the efficiency of the kinematic trigger selection has been
derived with data from control triggers, as described in Section~\ref{sec:trigger}, and is applied as a weight for
each event. Correction factors to account for the different \cPqb\ tagging efficiencies in data and simulation~\cite{BTV-16-002}
are also applied. The total signal efficiency ranges between 0.5 and 1.4\% and peaks around 500\GeV.
The efficiency first increases due to the kinematic selection and then decreases for masses beyond 500\GeV due to the requirement of three \cPqb-tagged jets, and the fact that the \cPqb\ tagging efficiency decreases at high jet \pt.

For nominal masses between 300 and 500\GeV, each signal shape is parameterized by a bifurcated Gaussian function, which has different widths on the right- and left-hand side of the peak position, continued at higher masses with an exponential function to describe the tail.
The function has five parameters. The signal of the 600\GeV mass
point requires one additional Gaussian function on each side of the peak position
to be able to describe the tails of the distribution. This function has nine
parameters in total.
For nominal masses in the range 700--1300\GeV, a Bukin function as defined in Appendix~\ref{sec:app:functions}, which has five parameters, is used.
All parameterizations provide a very good modelling of the \mjj spectra.

\begin{figure}[htpb]
	\centering
	\includegraphics[width=0.60\textwidth]{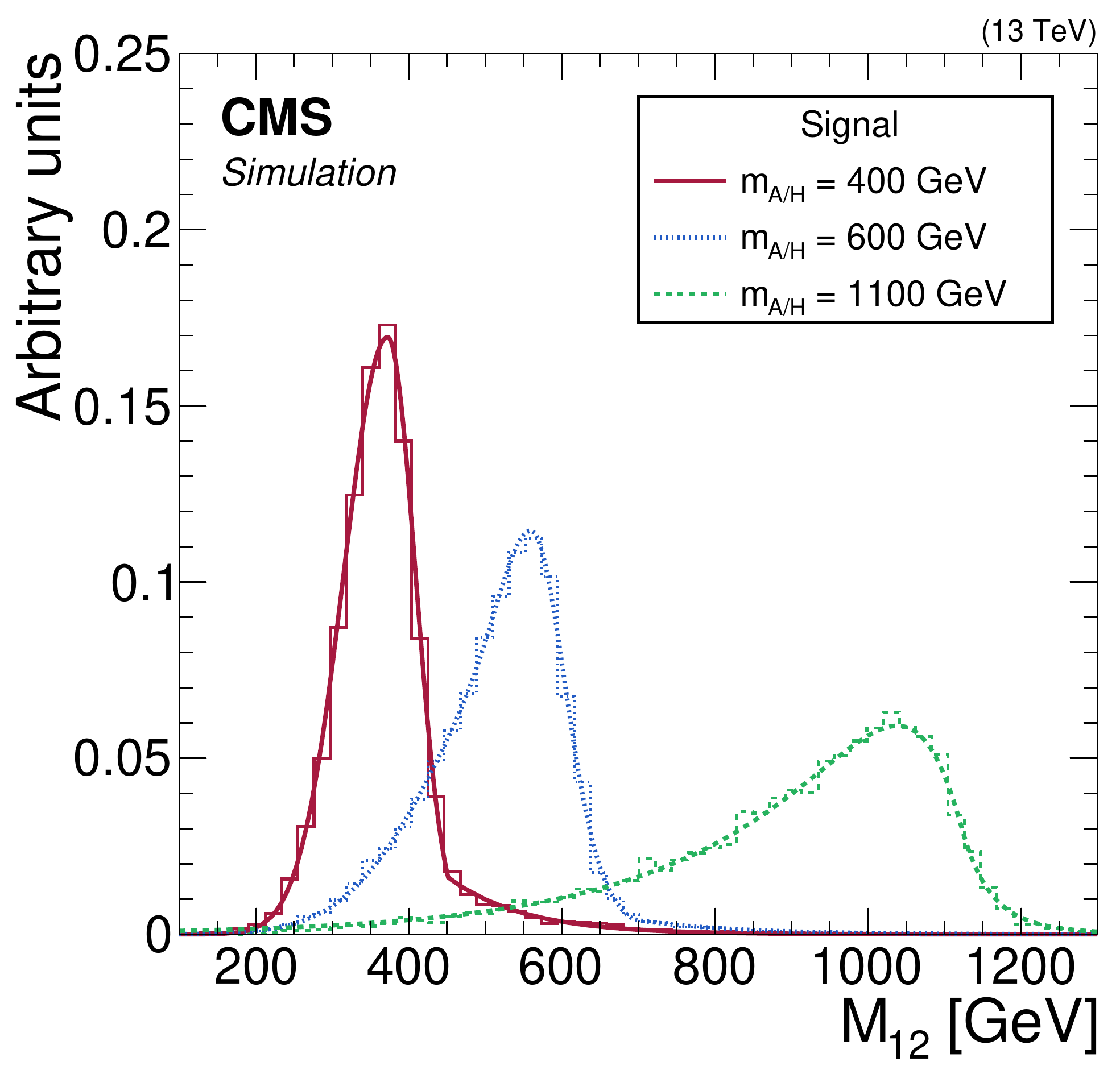}
	\caption{Invariant mass distributions of the two leading \cPqb\ jets in simulated signal events and their parameterizations
		for three different \PSA/\PH masses, normalized to unity.
	}
	\label{fig:signalHTemplates}
\end{figure}

The distributions of the invariant mass of the two leading \cPqb\ jets, \mjj, of the signal templates and parameterizations of the probability density function for different Higgs boson masses are shown in Fig.~\ref{fig:signalHTemplates}. The natural width expected for an MSSM Higgs boson in the considered mass and \tanb region is negligible compared to the detector resolution. For example, in the $\mhmodp$ scenario at a mass of 600\GeV and $\tanb = 60$, the natural width of the mass peak is found to be only about 19\% of the full width at half maximum of the reconstructed mass distribution. The shape of the mass distribution is thus dominated by the experimental resolution, and the possibility of the two leading jets used to compute \mjj not being the daughters of the Higgs boson, which we refer to as wrong jet pairing. Pronounced tails towards lower masses are attributed to cases of incomplete reconstruction of the Higgs daughter partons, for example due to the missing momentum of neutrinos in semileptonic decays of hadrons containing bottom and charm quarks. The wrong jet pairing gives rise to tails in both directions. For the lower mass points, however, the tails towards lower masses are suppressed because of the jet \pt threshold.

\section{Background model}
\label{sec:bkgdmod}
The main background for this analysis originates from multijet
production, with at least two energetic jets containing
\cPqb\ hadrons, and a third jet that satisfies the \cPqb\ tagging selection but possibly
as a result of a mistag. Top quark-antiquark production exhibits a shape very similar to the multijet process. It is found to be negligible, but nevertheless is implicitly covered by our background model.

The relevant features of the multijet background are studied in a
suitable control region (CR) in data, which is obtained from the triple \cPqb\ tag
selection by imposing a b-tag veto on the third
leading jet. This veto rejects jets that would satisfy a loose \cPqb\ tagging requirement,
defined by a 10\% probability for light-flavour jets to be misidentified
as \cPqb\ jets, and has a \cPqb\ jet identification efficiency of about 80\%. This CR has no overlaps with
the triple \cPqb\ tag signal region (SR), while it preserves similar kinematic
distributions for the three leading jets. In addition, the signal
contamination in the CR is negligible.

A suitably chosen analytic function is used to model the multijet
background. This function is extensively validated in the
\cPqb\ tag veto CR. In order to improve the background description and reduce the
potential bias related
to the choice of the background model, the \mjj distribution is divided into the three overlapping subranges [200, 650], [350, 1190], and [500, 1700]\GeV. Their borders are chosen to largely cover the signal shapes of the associated mass points of [300, 500], [500, 1100], and [1100, 1300]\GeV, respectively (as discussed in Section~\ref{sec:signaltempl}).

In the first subrange, the selection criteria introduce a kinematic edge (turn-on) in the \mjj distribution. The chosen function is a product of two terms. The first term is a turn-on function, represented by a Gaussian error function in the form of:
\begin{equation}
f(\mjj) = 0.5\, \bigl[\erf(p_0[\mjj - p_1])+1\bigr],
\end{equation}
where
\begin{equation}
\erf(x) = \frac{2}{\sqrt{\pi}} \int_0^{x} \re^{-t^2} \rd{}t,
\end{equation}
and the parameters $p_0$ and $p_1$ describe the slope and point of the turn-on, respectively.

The falling part of the spectrum is described by an extension of the Novosibirsk function originally used to describe a Compton spectrum~\cite{Ikeda:1999aq}, defined as:
\begin{equation}
g(\mjj) = p_2 \exp \left( - \frac{1}{2\sigma_{0}^{2}} \ln^{2}[ 1 - \frac{\mjj - p_3}{p_4} p_5 -  \frac{(\mjj - p_3)^2}{p_4} p_5 p_6] - \frac{\sigma_0^2}{2} \right),
\end{equation}
where $p_2$ is a normalization parameter, $p_3$ the peak value of the distribution, $p_4$ and $p_5$ are the parameters describing the asymmetry of the spectrum, and $p_6$ is the parameter of the extended term.
The variable $\sigma_0$ is defined as:
\begin{equation}
\sigma_0 = \frac{2}{\xi} \sinh^{-1}(p_5 \xi / 2),\ \text{where}\ \xi = 2 \sqrt{\ln 4}.
\end{equation}
In the second and third subranges, we choose a nonextended Novosibirsk function ($p_6 \equiv 0$)
without turn-on factor.

\begin{figure}[htpb]
	\centering
	\includegraphics[width=0.48\textwidth]{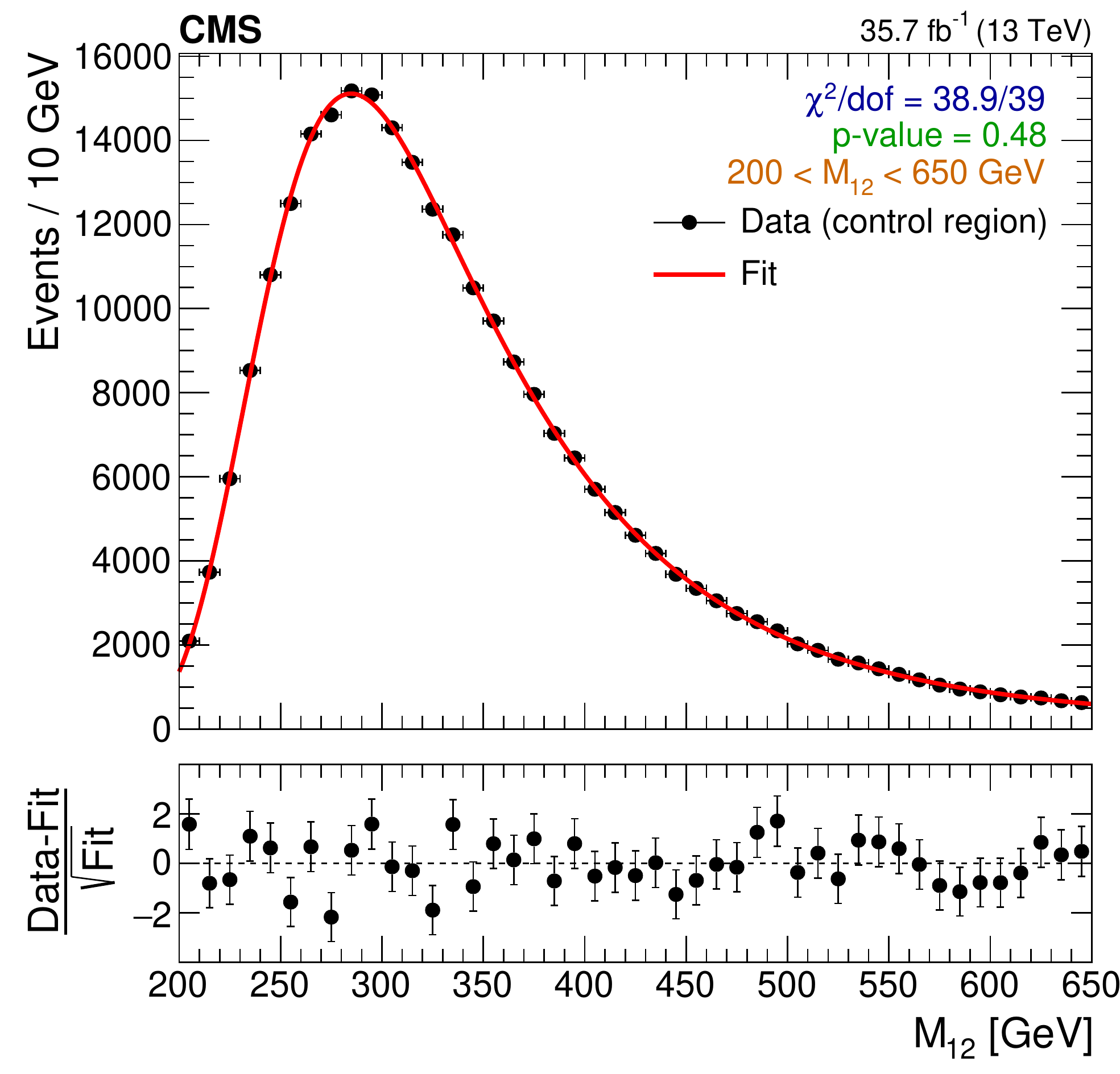}\hfill
	\includegraphics[width=0.48\textwidth]{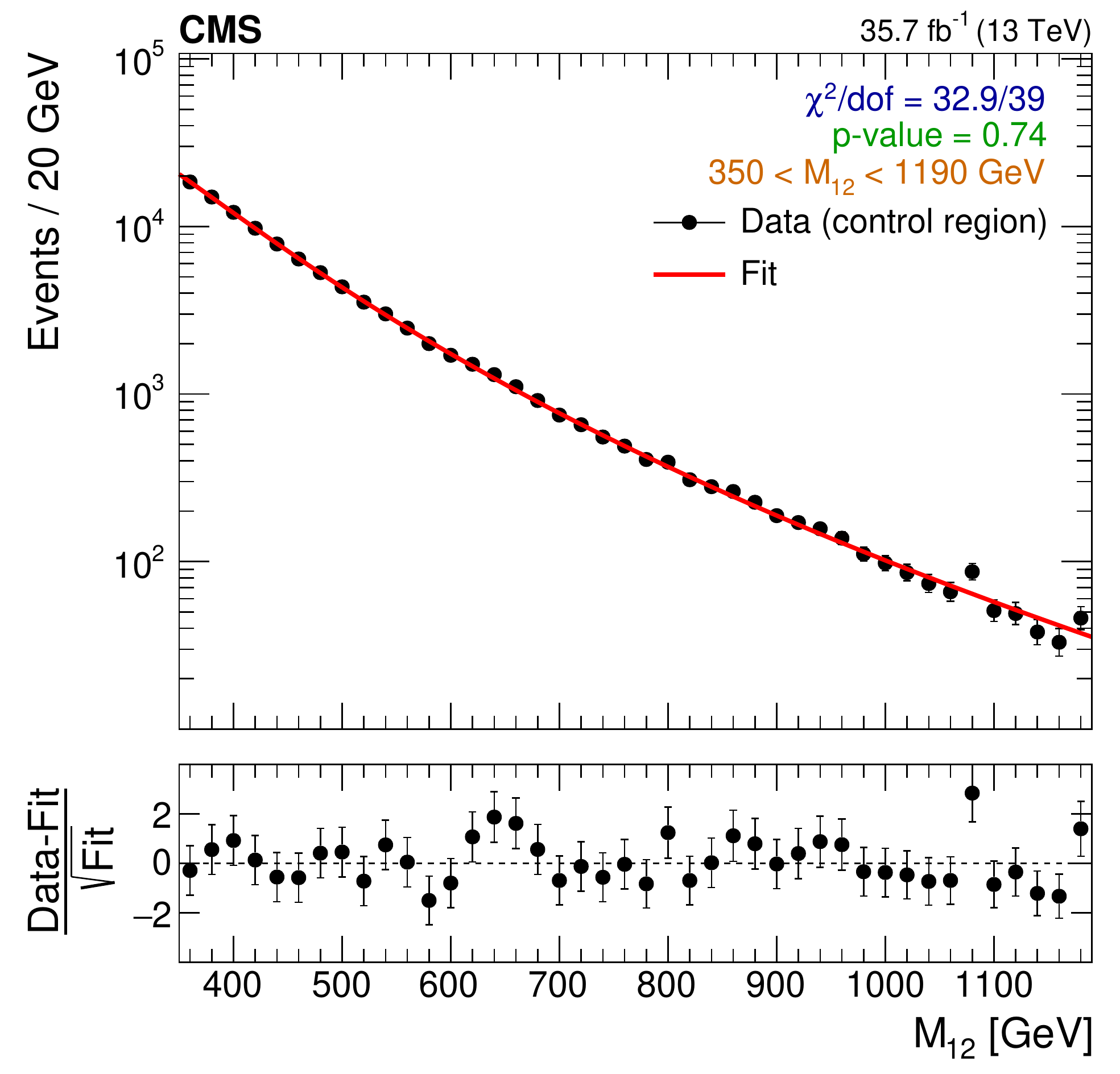}\\
	\includegraphics[width=0.48\textwidth]{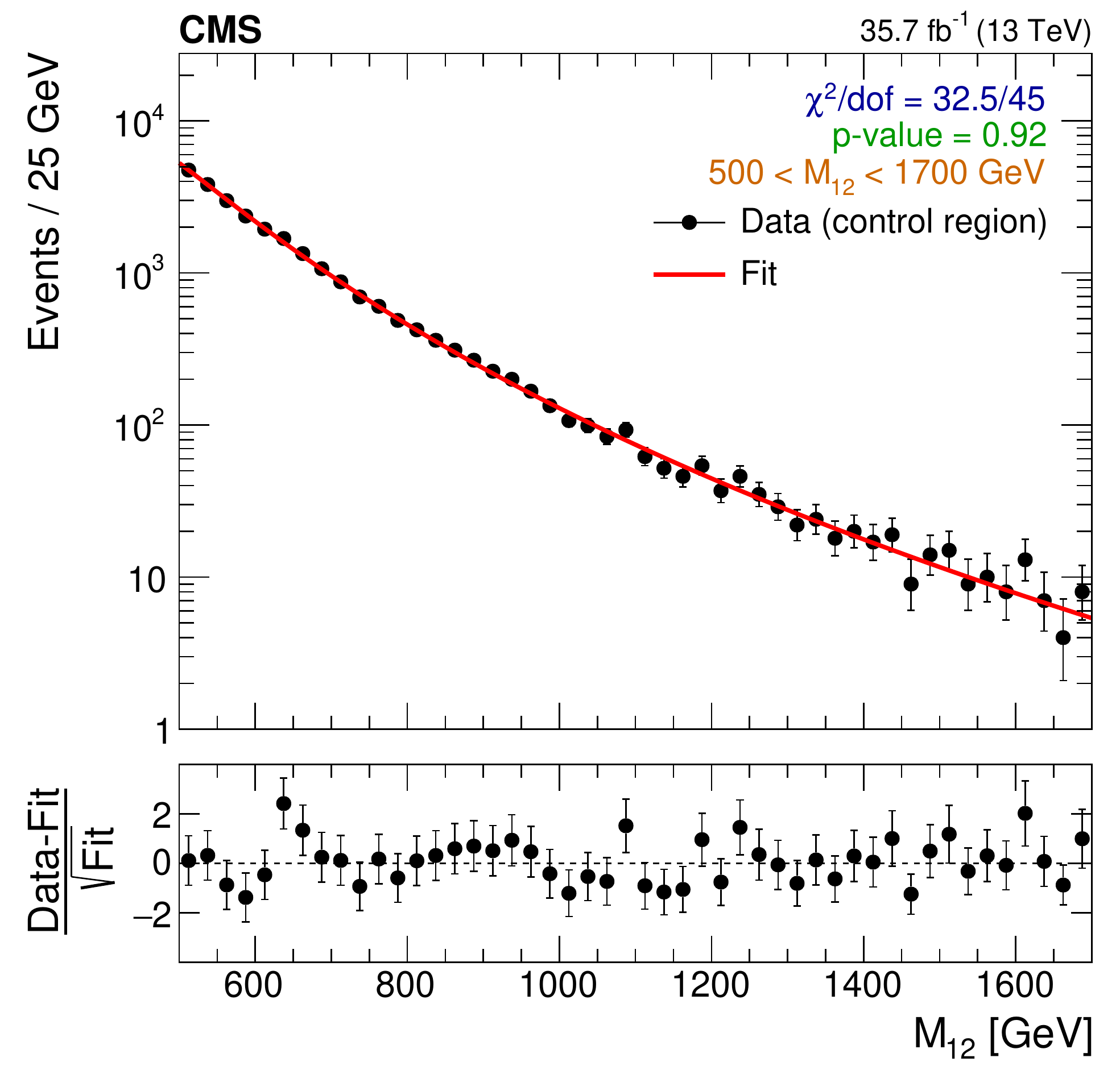}\\
	\caption{Distributions of the dijet invariant mass \mjj, obtained from the \cPqb\ tag veto CR as described in the text in the three subranges
		used for the fit: $\mjj = [200, 650]\GeV$ (upper left) in linear scale,  $\mjj = [350, 1190]\GeV$ (upper right) and $\mjj = [500, 1700]\GeV$ (lower) in logarithmic scale. The dots represent the data. The full line is the result of the fit of the background parameterizations described in the text. In the bottom panel of each plot, the normalized difference [(Data-Fit)/$\sqrt{\mathrm{Fit}}$] is shown.}
	\label{fig:templatesHighMass}
\end{figure}

Figure~\ref{fig:templatesHighMass} shows the fits of the chosen functions to the CR data, which have been prescaled to give similar event count as in the SR. In the first subrange,
$\mjj = [200, 650]\GeV$, the turn-on effect due to the jet \pt threshold at trigger level is
clearly visible. In the other two mass subranges, the spectrum shows only the expected falling behaviour
with \mjj. The values of the parameters $p_0$ and $p_1$ used to model the turn-on obtained in the CR are also used for the SR fit since the turn-on behaviour in the two regions is found to be very similar. The other
function parameters are allowed to vary independently in the CR and SR fits.

Different families of alternative probability density functions such as Bernstein polynomials and the so-called dijet function as defined in Ref.~\cite{Sirunyan:2016cao} are studied to estimate the possible bias from the choice of the background model.
For each family, a systematic bias on the extraction of a signal with mass \mAH is determined: the alternative function is fit to the observed data, from which toy experiments are drawn. Using the nominal background model in the respective subrange, a maximum-likelihood fit of signal and background is performed for each pseudo-experiment.
The difference in the extracted and injected number of signal events is divided by the statistical uncertainty of the fit.
The resulting pull distribution is considered to represent the systematic bias on the signal strength
due to the choice of the background function and our insufficient knowledge of the background processes. We infer a bias of 100, 20, and 25\% in units of the statistical uncertainty of the signal strength for the first, second, and third subranges, respectively.

\section{Systematic uncertainties}
\label{sec:Systematics}

The following systematic uncertainties in the expected signal and
background estimation affect the determination of the signal yield or its
interpretation within the MSSM or generic 2HDM models.

The signal yields are affected by the following uncertainties:
\begin{itemize}
	\item a 2.5\% uncertainty in the estimated integrated luminosity of the data sample~\cite{CMS-PAS-LUM-17-001};
	\item the uncertainty in the online \cPqb\ tagging efficiency scale factor, which results in an
	overall uncertainty in the range of 0.8--1.3\% for Higgs boson
	masses of 300--1300~GeV;
	\item a 5\% uncertainty in the correction of the selection efficiency comparing to the NLO prediction;
	\item the effect due to the choice of PDFs and the value of $\alpha_s$ (1--6\%), following the recommendations of the LHC Higgs Cross Section Working Group~\cite{deFlorian:2016spz} when interpreting the results in benchmark models;
	\item the uncertainty in the normalization and factorization scales (1--10\%) when interpreting the results in benchmark models.
\end{itemize}

Uncertainties affecting the shape as well as the normalization of the signal
templates are:
\begin{itemize}
	\item the uncertainty in the jet trigger efficiencies, ranging between subpercent values and 7\% per jet depending on its $\eta$ and \pt;
	\item the uncertainty in the off\/line \cPqb\ tagging efficiency (2--5\% per jet depending on its transverse momentum) and the mistag scale factors ($<$0.3\%);
	\item the jet energy scale (JES) and jet energy resolution (JER) uncertainties (1--6\%):
	their impact is estimated by varying the JES and JER in the simulation within the measured
	uncertainties;
	\item the uncertainty in the total inelastic cross section of 4.6\% assumed in the
	pileup simulation procedure~\cite{Sirunyan:2017hlu}.
\end{itemize}

For the background estimation, the bias on the extracted signal strength,
as reported in Section~\ref{sec:bkgdmod}, is considered as an additional bias term to the background fitting function. This poses the largest uncertainty for the analysis.

\section{Results}
\label{sec:results}

\begin{figure}[htbp]
	\centering
	\includegraphics[width=1.0\textwidth]{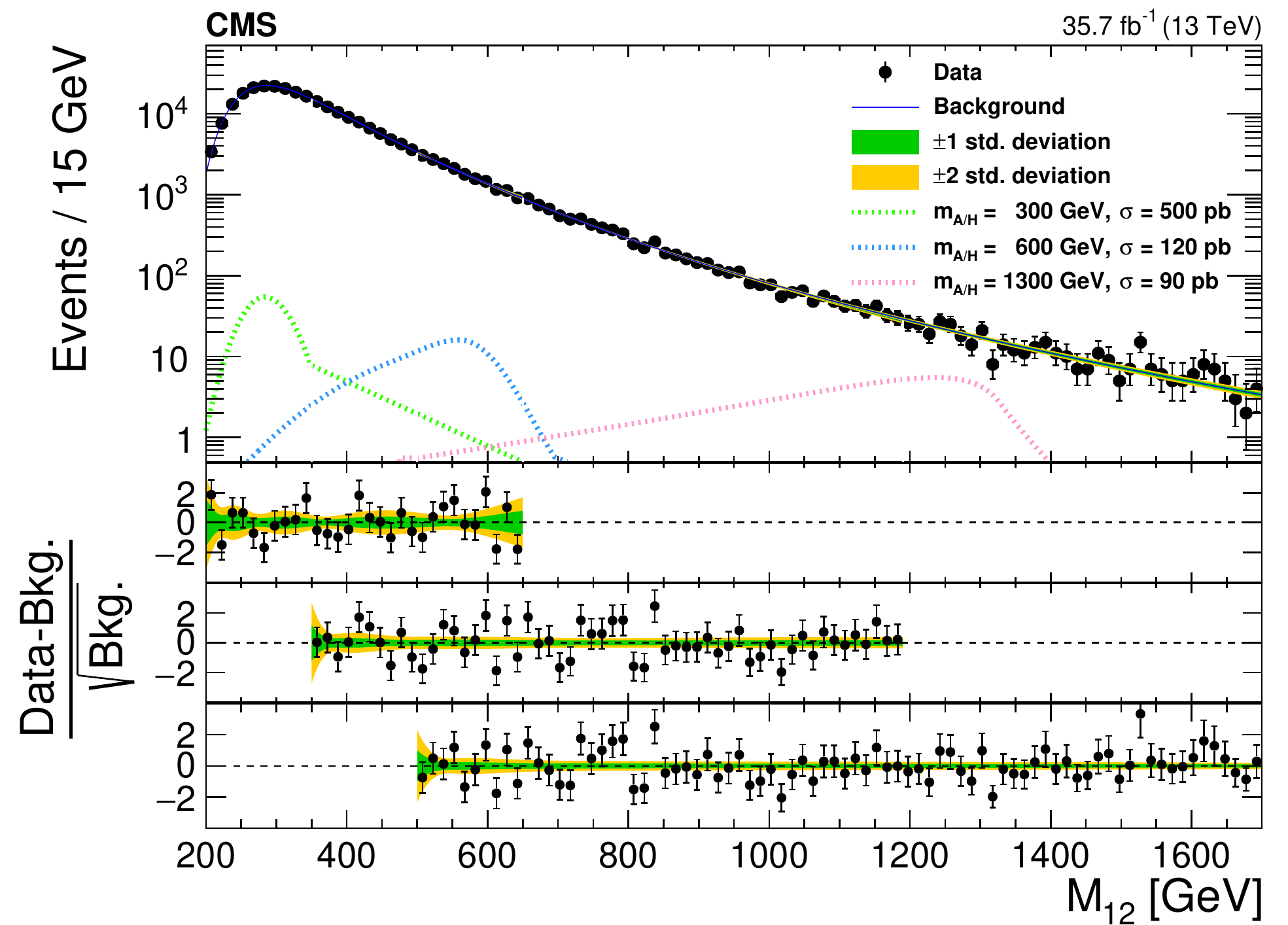}\hfill
	\caption{Distribution of the dijet invariant mass \mjj in the data
		triple \cPqb\ tag sample showing the three subranges together
		with the corresponding background-only fits.
		The shaded area shows the post-fit uncertainty.
		For illustration, the expected
		signal contribution for three representative mass points is
		shown, scaled to cross sections suitable for visualization.
		The change of slope around 350\GeV of the 300\GeV signal shape is
		caused by wrong jet pairing.
		In the bottom panels the normalized difference
		((Data-Bkg)/$\sqrt{\mathrm{Bkg}}$),
		where Bkg is the background as estimated by the fit,
		for the three subranges is shown.  }
	\label{fig:template_fit_allInOne}
\end{figure}

The number of potential signal events is extracted by performing a maximum-likelihood fit of the
signal plus background parameterizations to the \mjj data distribution.
Initially, a fit with only the background parameterizations is performed.
Results of this background-only fit in all three subranges are given in
Fig.~\ref{fig:template_fit_allInOne}. A good description of the data is observed.
The normalized differences between data
and fit together with the post-fit uncertainties are shown for each subrange.

{\tolerance=5000
In a second step, a combined fit of signal and background to the data is performed.
No significant excess over the background-only distribution is observed and upper limits
at 95\% confidence level (\CL) on the cross section times branching fraction
${\sigma(\Pp\Pp\to\PQb \PSA/\PH+\mathrm{X}) \mathcal{B}(\PSA/\PH\to\bbbar)}$ are derived.
For the calculation of exclusion limits, the modified frequentist
criterion \CLs~\cite{Junk:1999kv,Read:2002hq,Cowan:2010js} is adopted using the \textsc{{\allowbreak Roo}{\allowbreak Stats}}
package~\cite{RooStats}.
The test statistic is based on the profile likelihood ratio. Systematic uncertainties are treated as nuisance parameters and profiled in the statistical interpretation using log-normal priors for uncertainties affecting the signal yield, while Gaussian priors are used for shape uncertainties. \par}

\begin{figure}[htbp]
	\centering
	\includegraphics[width=0.7\textwidth]{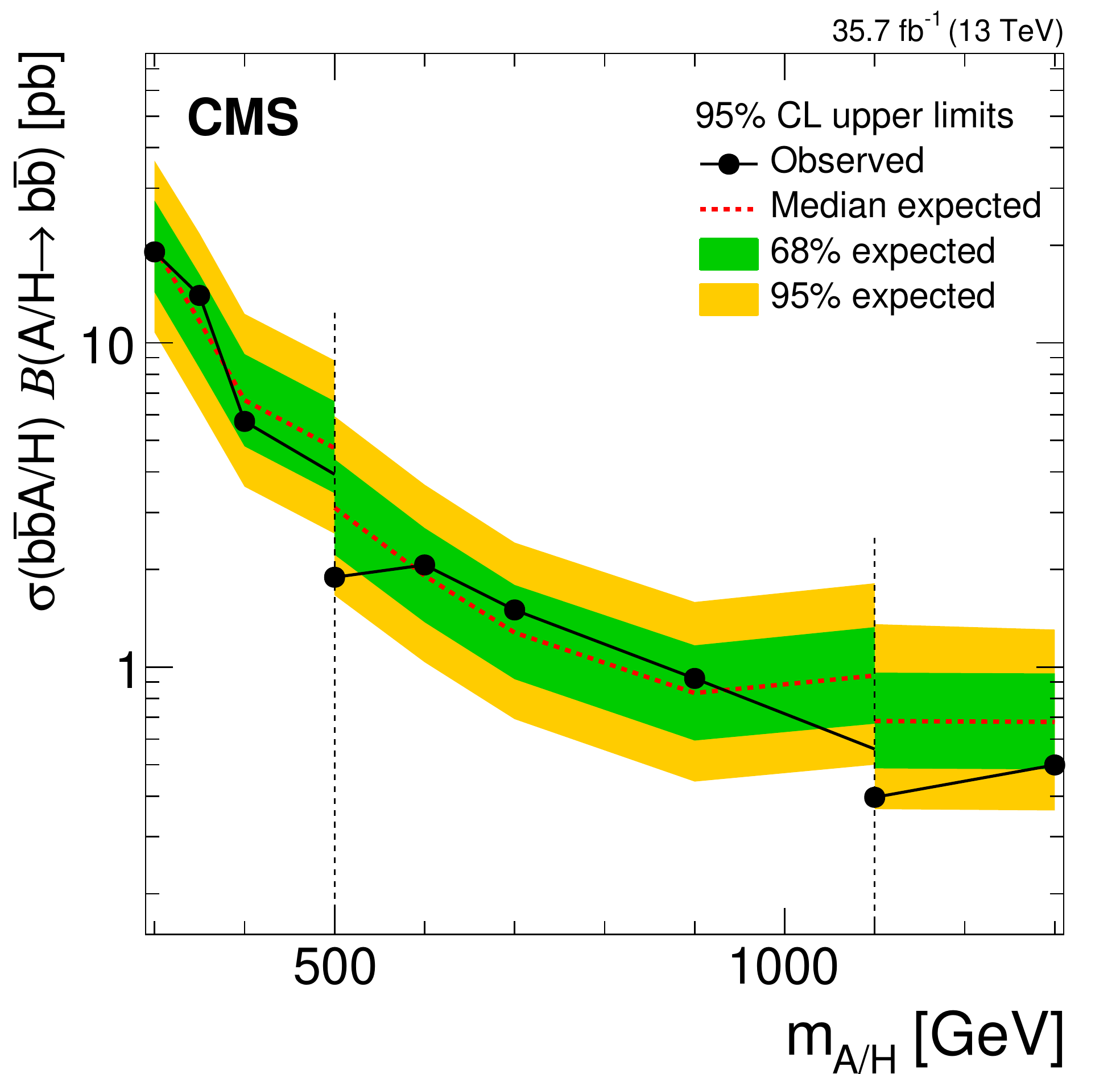}
	\caption{\label{fig:xsec_mA_bands_csvt} Expected and observed upper limits
		on $\sigma(\Pp\Pp\to\PQb \PSA/\PH+\mathrm{X})\,\mathcal{B}(\PSA/\PH\to\bbbar)$
		at \clsnf\ as a function of the Higgs boson mass \mAH.
		The inner and the outer bands indicate the regions containing 68 and 95\%, respectively, of the distribution of limits expected under the background-only hypothesis.
		The dashed horizontal lines illustrate the borders between the three subranges in which the results have been obtained.}
	
\end{figure}

Model-independent upper cross section times branching fraction limits are shown as a function of the mass of the \PSA/\PH bosons in Fig.~\ref{fig:xsec_mA_bands_csvt} up to a mass of 1300\GeV. The visible steps in the expected and observed limits at 500 and 1100\GeV are due to the transitions between the mass subranges as explained in Section~\ref{sec:bkgdmod}. The limits range from about 20\unit{pb} at 300\GeV, to about 0.4\unit{pb} at 1100\GeV. The limits are also summarized in Table~\ref{tab:limits:xsec} in Appendix~\ref{sec:app:limits}.

\subsection{Interpretation within the MSSM}
The cross section limits shown in Fig.~\ref{fig:xsec_mA_bands_csvt} are
translated into exclusion limits on the MSSM parameters
\tanb and \mA. The cross sections for $\PQb +  \PSA/\PH$ associated production
as obtained with the four-flavour NLO~\cite{Dittmaier:2003ej, Dawson:2003kb} and the
five-flavour NNLO QCD calculations implemented in
\textsc{bbh@nnlo}~\cite{Harlander:2003ai} were combined using the Santander matching
scheme~\cite{Harlander:2011aa}. The
branching fractions were computed with
\textsc{FeynHiggs} version 2.12.0~\cite{Degrassi:2002fi,Frank:2006yh,Heinemeyer:1998yj,Heinemeyer:1998np}
and \textsc{Hdecay}~\cite{Djouadi:1997yw,Djouadi:2006bz} as described in Ref.~\cite{Heinemeyer:2013tqa}.

\begin{figure}[htbp]
	\centering
	\includegraphics[width=0.48\textwidth]{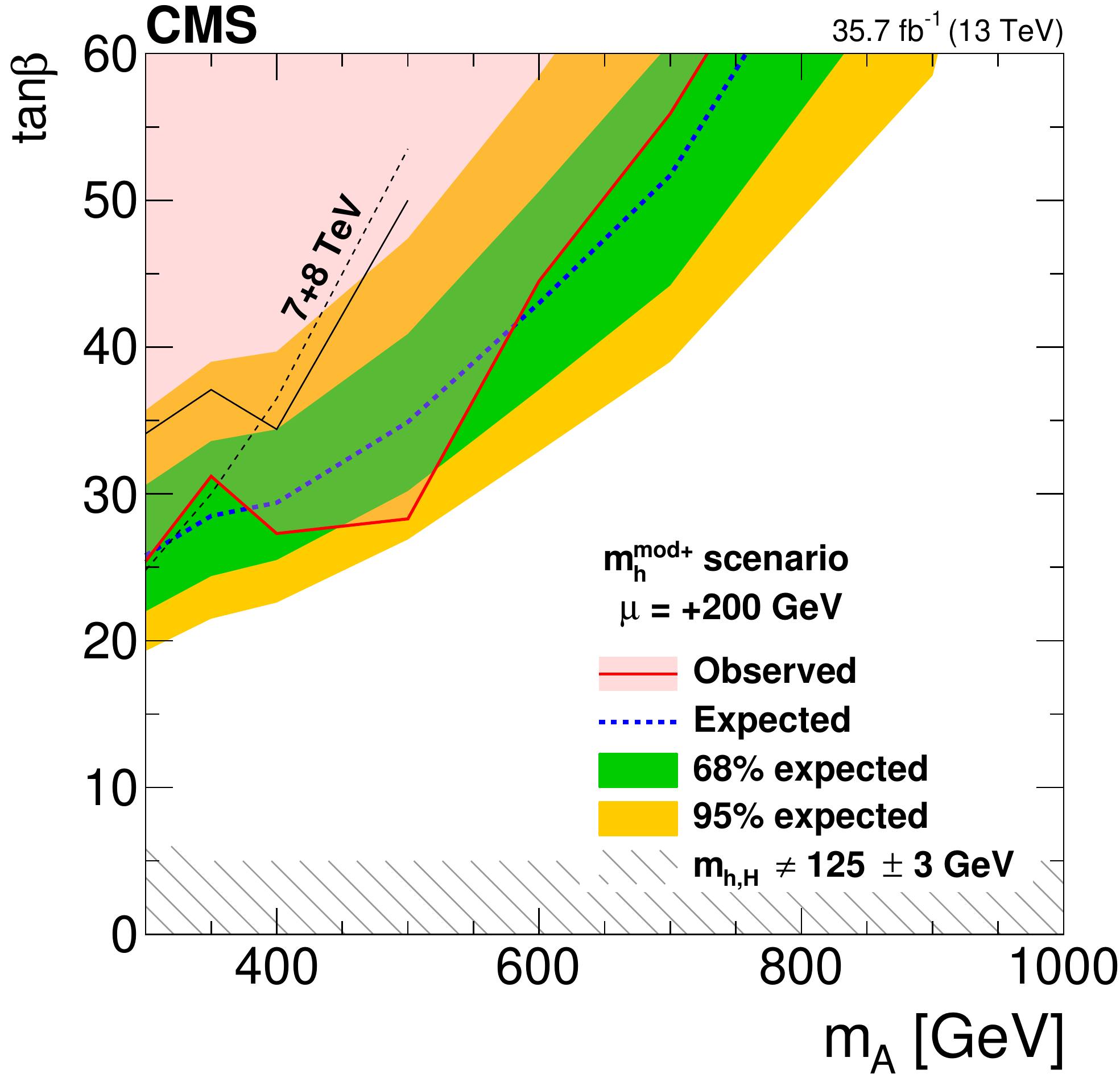}\hfill
	\includegraphics[width=0.48\textwidth]{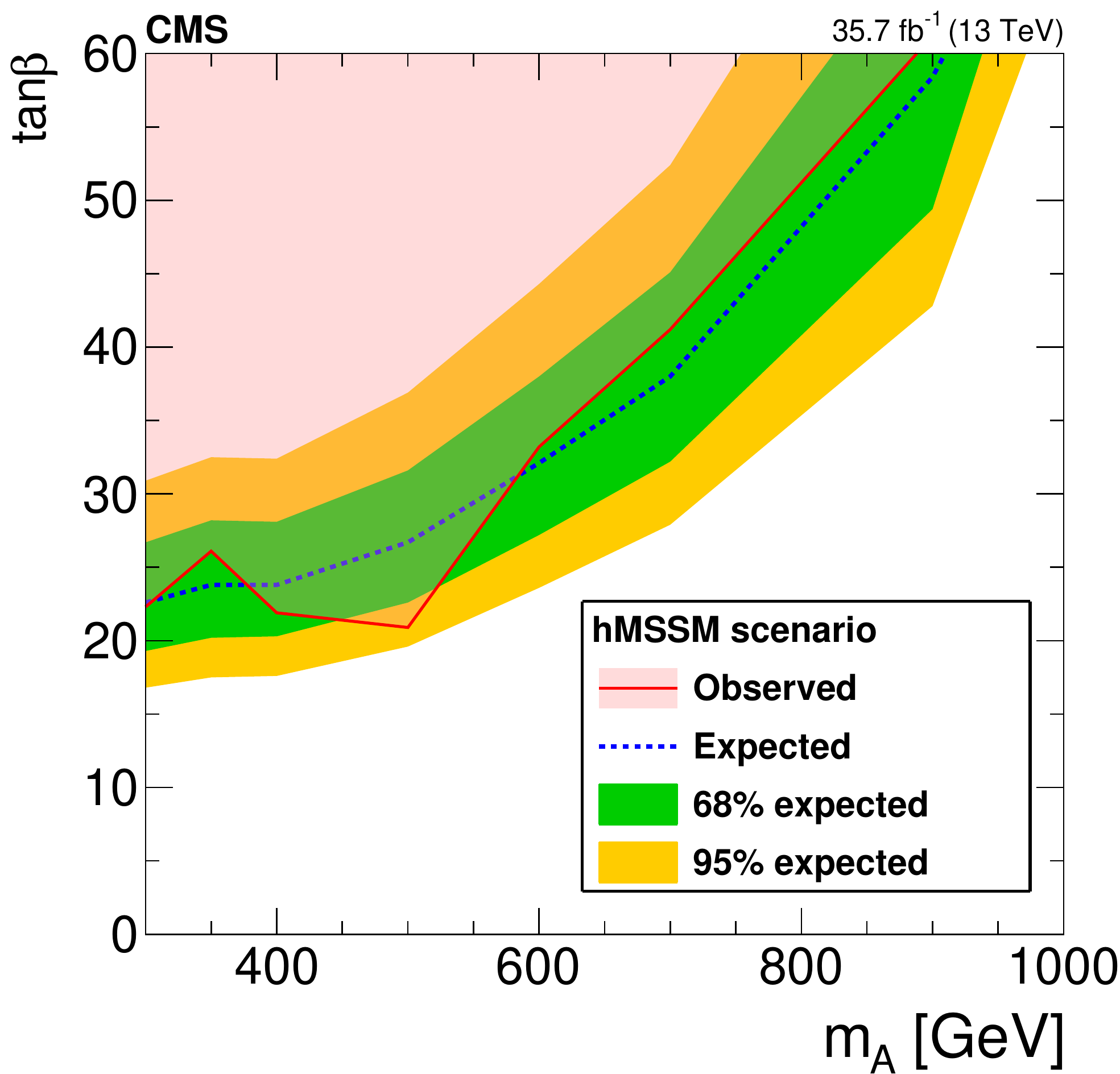}\hfill
	\includegraphics[width=0.48\textwidth]{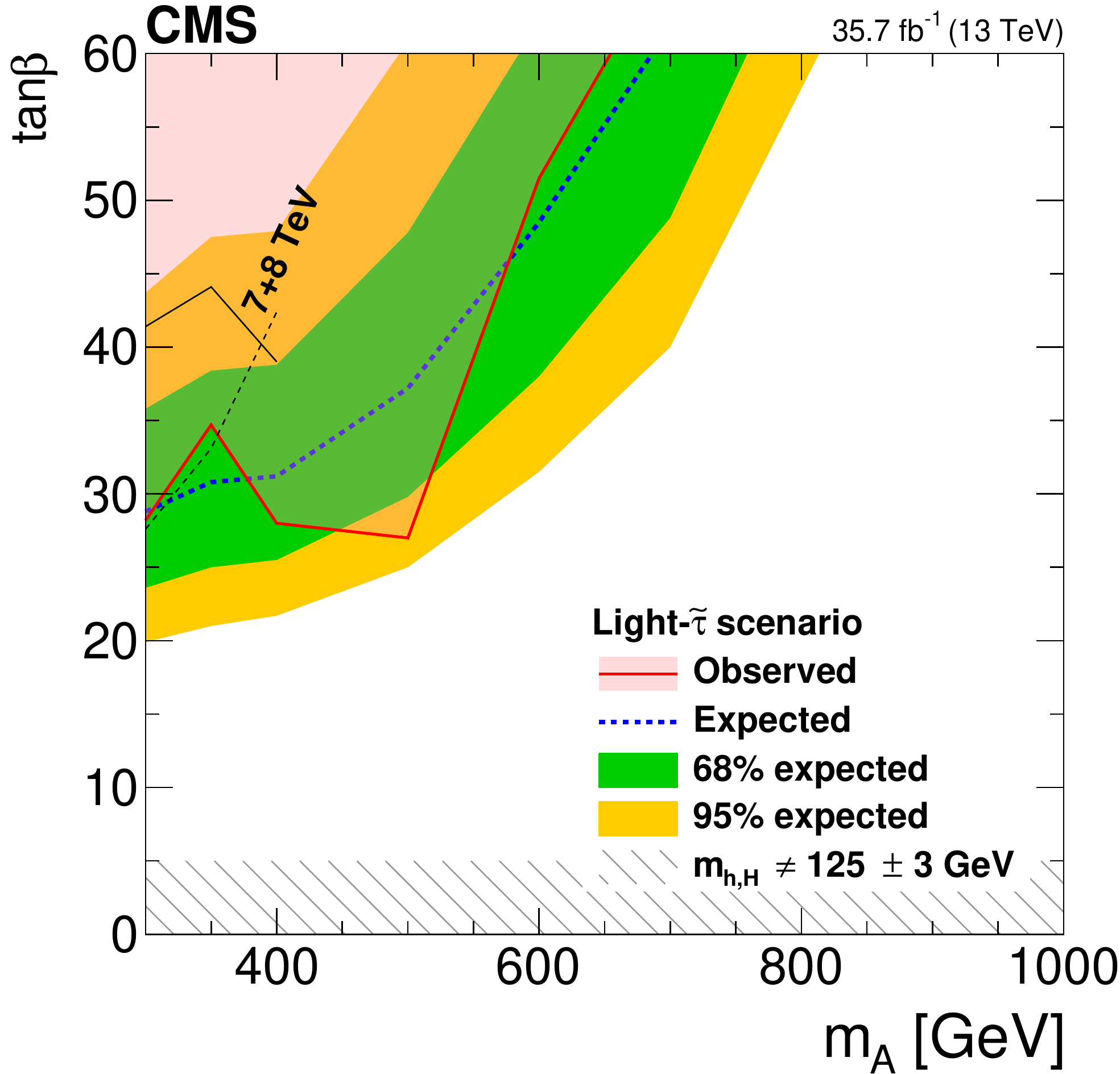}\hfill
	\includegraphics[width=0.48\textwidth]{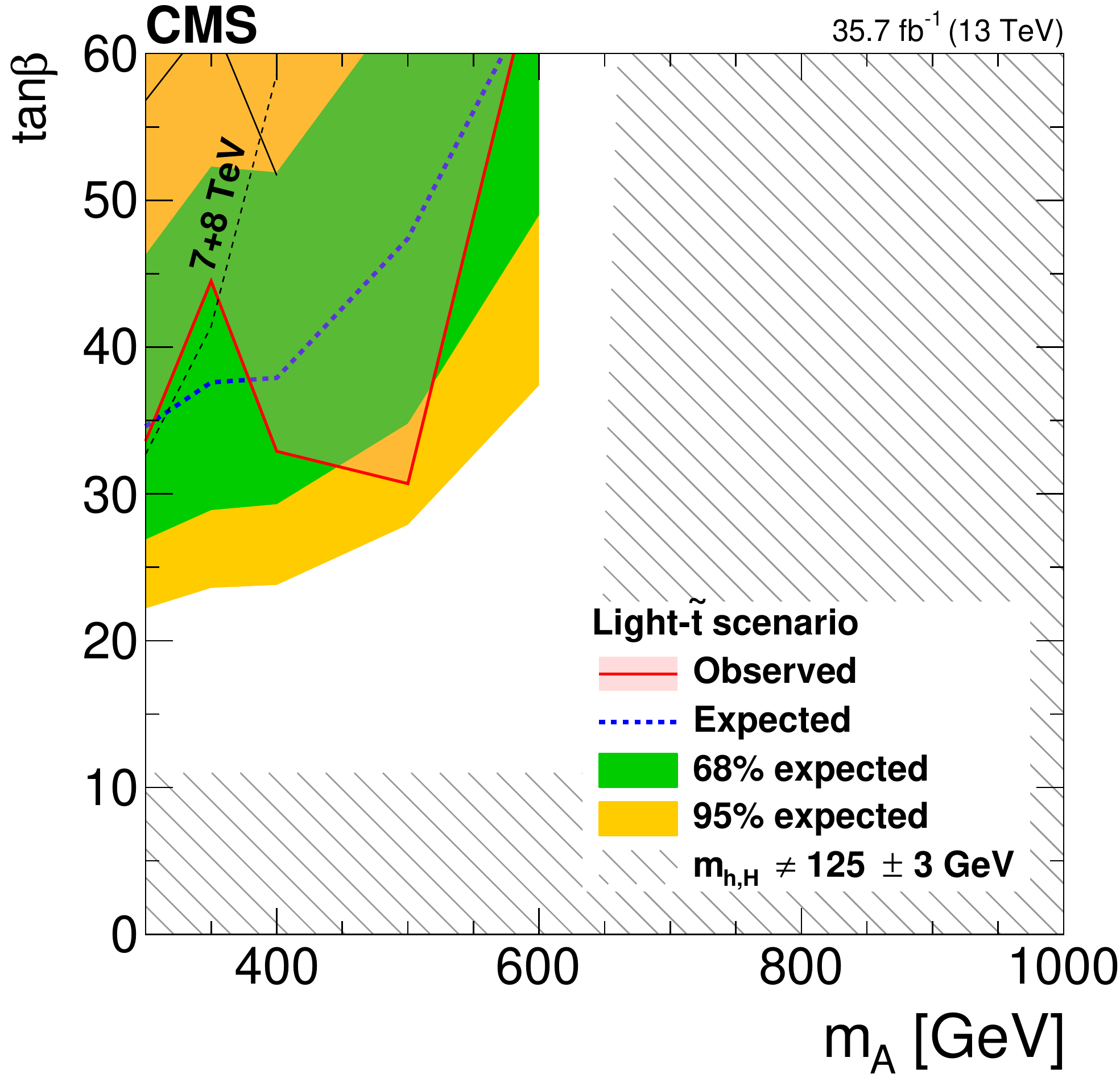}\hfill
	
	\caption{Expected and observed upper limits at \clsnf\ for \mA vs.\
		the MSSM parameter \tanb in the (upper left) $m_\Ph^\mathrm{mod+}$
		benchmark scenario with $\mu=+200\GeV$, in the (upper right) hMSSM,
		the (lower left) light $\sTau$, and the (lower right) light $\sTop$ benchmark scenarios.
		The inner and outer bands indicate the regions containing 68 and 95\%, respectively, of the distribution of limits expected under the background-only hypothesis.
		The excluded parameter space is indicated by the red shaded area. The hashed area is excluded because $m_{\Ph,\PH}$ would deviate by more than $\pm 3\GeV$ from the mass of the observed Higgs boson at 125\GeV. Since theoretical calculations for $\tanb > 60$ are not reliable, no limits are set beyond this value.
To illustrate the improvement in sensitivity, the observed and expected upper limits from the preceding CMS analyses at
$\sqrt{s} = 7$ and 8\TeV~\cite{Chatrchyan:2013qga,Khachatryan:2015tra} are also shown as solid and dashed black lines. 
	}
	\label{fig:tanb_mA}
\end{figure}

The observed and expected \clsnf\ median upper limits on \tanb
versus \mA are shown in Fig.~\ref{fig:tanb_mA} (upper row). They were
computed within the
MSSM $m_\Ph^\mathrm{mod+}$ benchmark scenario~\cite{Carena:2005ek} with the higgsino mass parameter $\mu = +200\GeV$ and in the hMSSM scenario~\cite{Maiani:2013hud,Djouadi:2013uqa,Djouadi:2015jea}.
In the former scenario, the observed
upper limits range from \tanb of about 25 at $\mA=300\GeV$ to about 60 at
$\mA=750\GeV$. These results considerably extend the preceding measurements at
$\sqrt{s} = 7$ and 8\TeV~\cite{Chatrchyan:2013qga,Khachatryan:2015tra}.
The model interpretation is not extended beyond \tanb values of 60, as theoretical predictions are not considered reliable for much higher values.
Additional model interpretations for \mA vs.\
\tanb in the light $\sTau$ and the light $\sTop$ benchmark scenarios are given in Fig.~\ref{fig:tanb_mA} (lower), and in Tables~\ref{tab:limits:mhmodp}--\ref{tab:limits:lightstop} in Appendix~\ref{sec:app:limits}.

\subsection{Interpretation within the 2HDM}

Cross sections and branching fractions for the \cPqb\cPqb\PH and \cPqb\cPqb\PSA\ processes
within different 2HDM models have been computed at NNLO using
\textsc{SusHi} version 1.6.1~\cite{Harlander:2012pb}, \textsc{2hdmc} version 1.7.0~\cite{Eriksson:2009ws} and
\textsc{lhapdf} version 6.1.6~\cite{Buckley:2014ana}. The 2HDM parameters have
been set according to the ``Scenario G'' proposed in Ref.~\cite{Haber:2015pua}.
Specifically, the heavier Higgs bosons are assumed to be degenerate in mass (\mA = \mH = $m_\PHpm$), and the mixing term
has been set to $m_{12}^2 = 0.5 \mA^2 \sin 2\beta$. The choice of such an MSSM-like parameterization allows
using the same signal samples as for the MSSM analysis.

\begin{figure}[htpb]
	\includegraphics[width=0.48\textwidth]{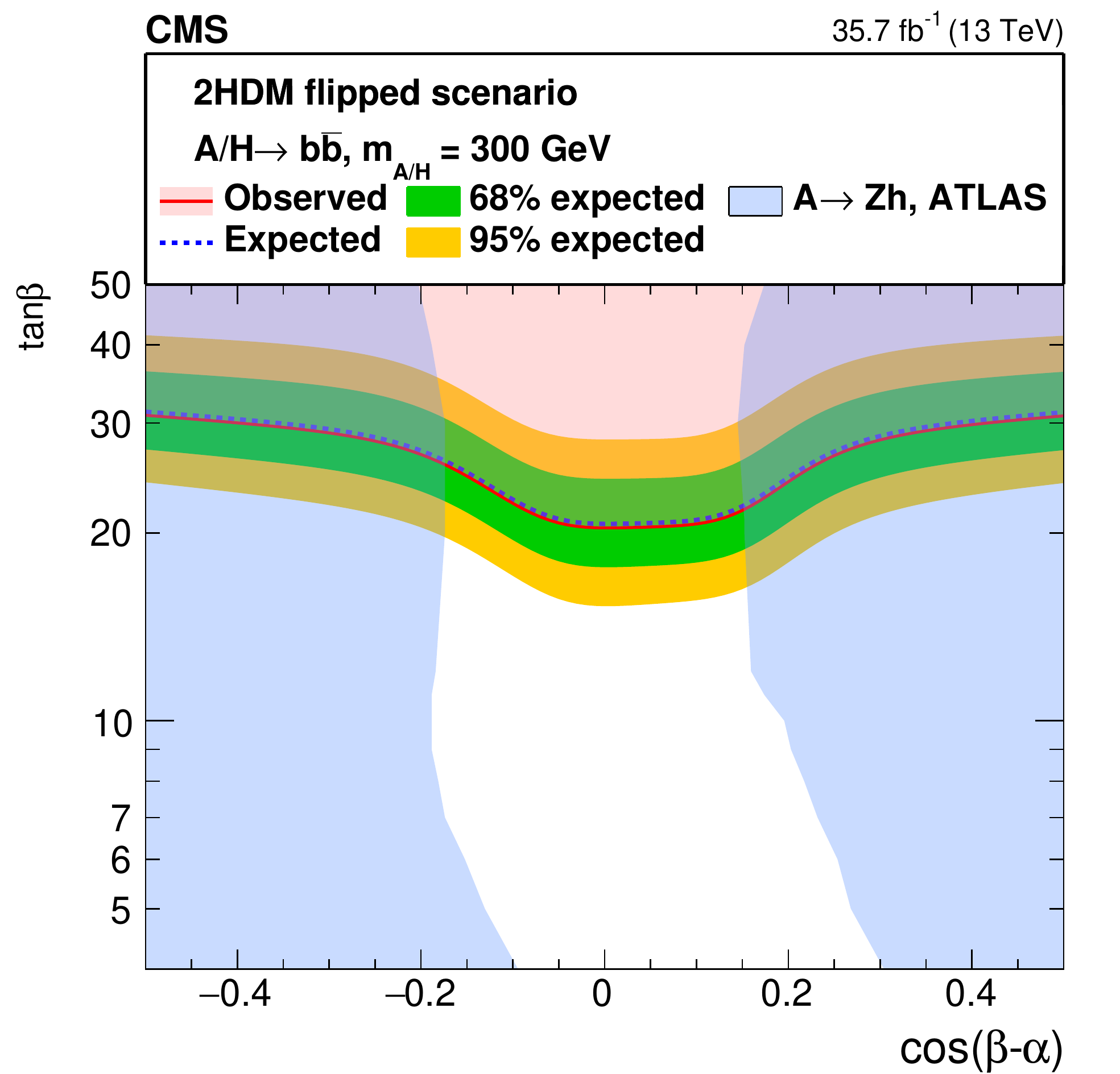}\hfill
	\includegraphics[width=0.48\textwidth]{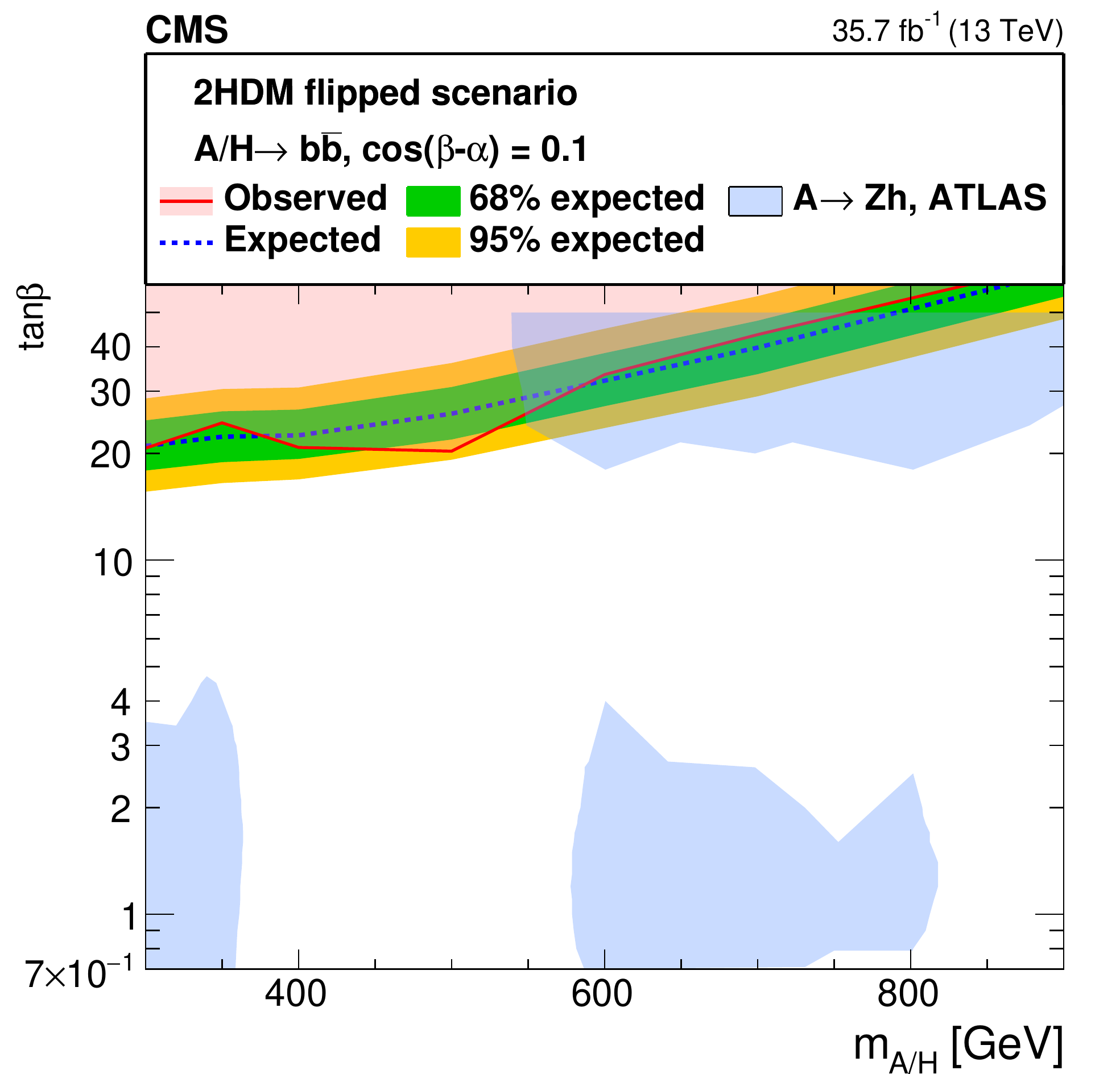}\\
	\includegraphics[width=0.48\textwidth]{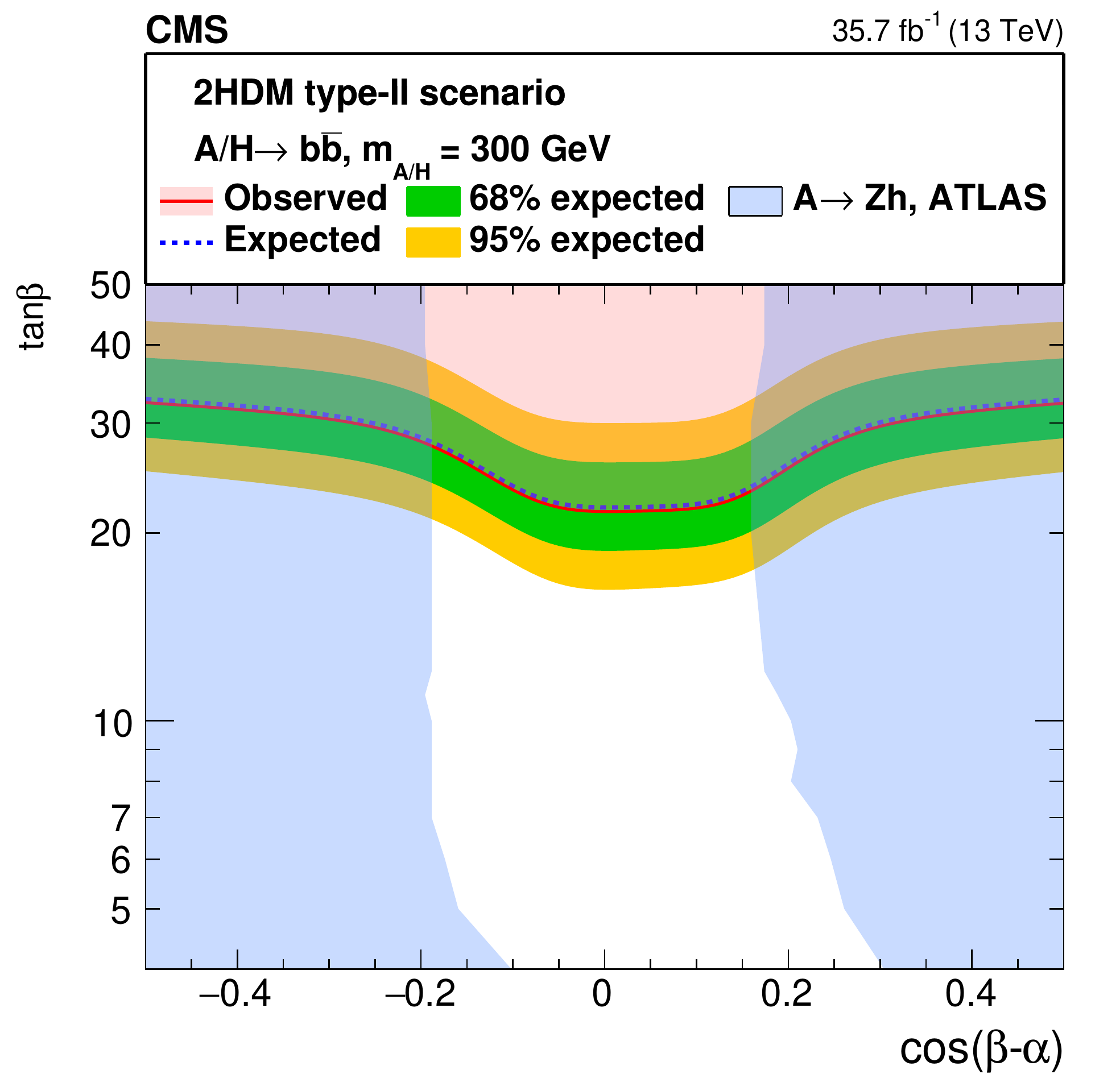}\hfill
	\includegraphics[width=0.48\textwidth]{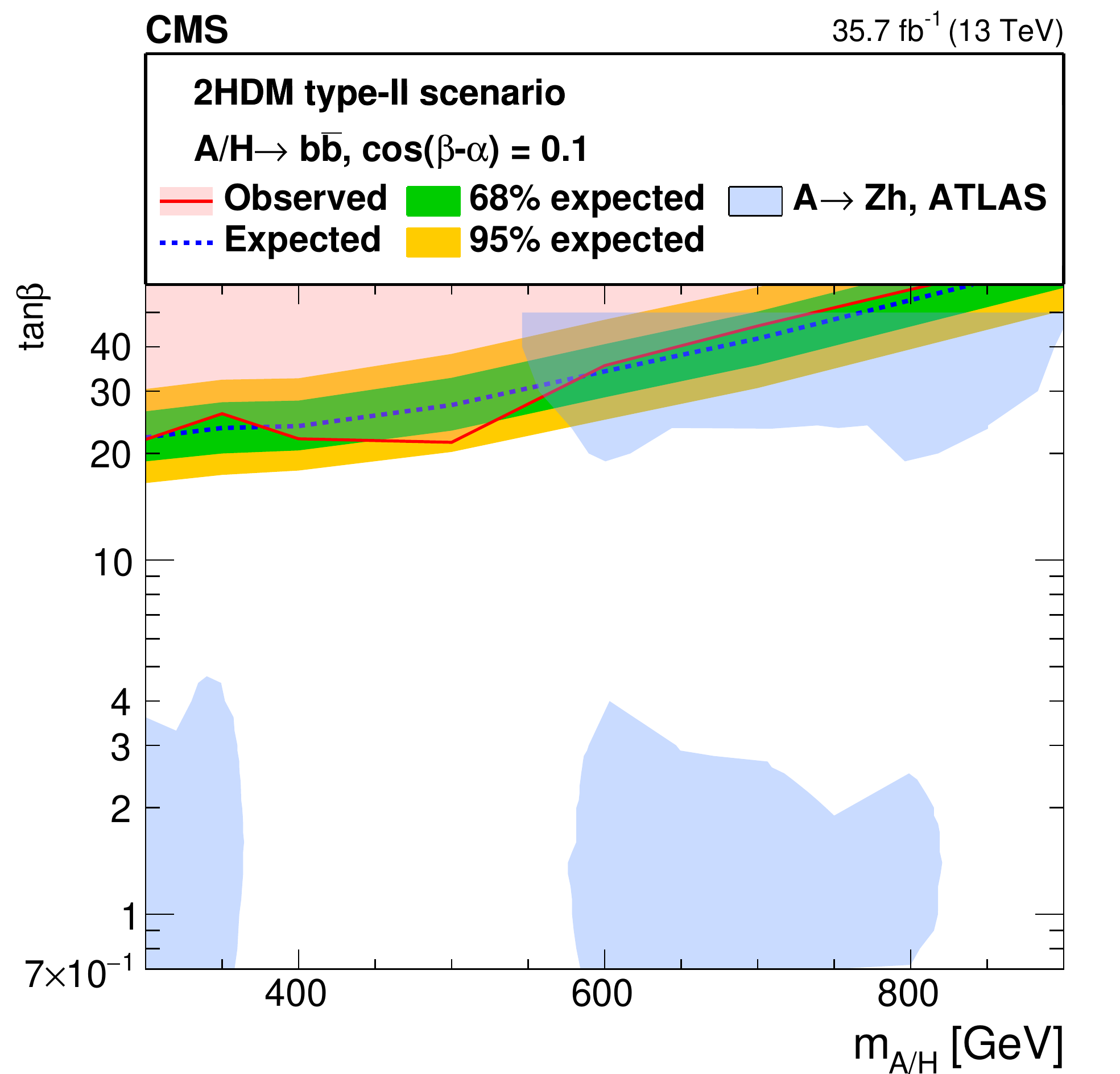}\\
	\caption{Upper limits for the parameter \tanb at 95\%
		\CL\ for the flipped (upper) and type-II (lower) models, as a function of \cosba
		in the range of $[-0.5,0.5]$ for the mass $m_\PH=m_\PSA=300\GeV$ (left) and as a function of \mAH
		when $\cosba = 0.1$ (right). The observed limits from the ATLAS $\PSA \to \PZ\Ph$
		analysis~\cite{Aaboud:2017sjh} at 95\% \CL, which are provided up to $\tanb = 50$, are also shown as blue shaded area for comparison.
	}
	\label{Fig:Interpretation:ULBrazil2HDM}
\end{figure}

\begin{figure}[htpb]
	\includegraphics[width=0.48\textwidth]{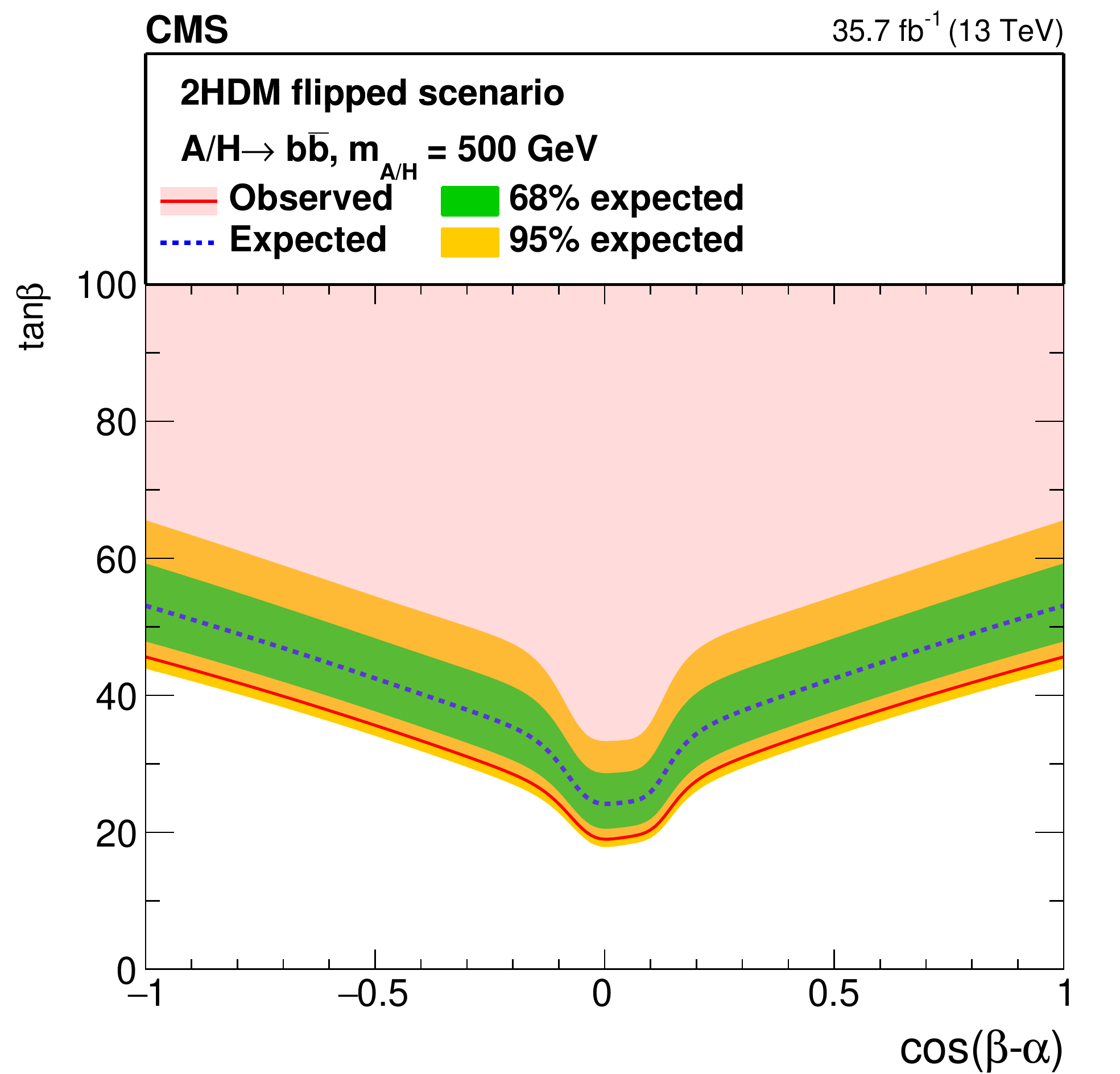}
	\includegraphics[width=0.48\textwidth]{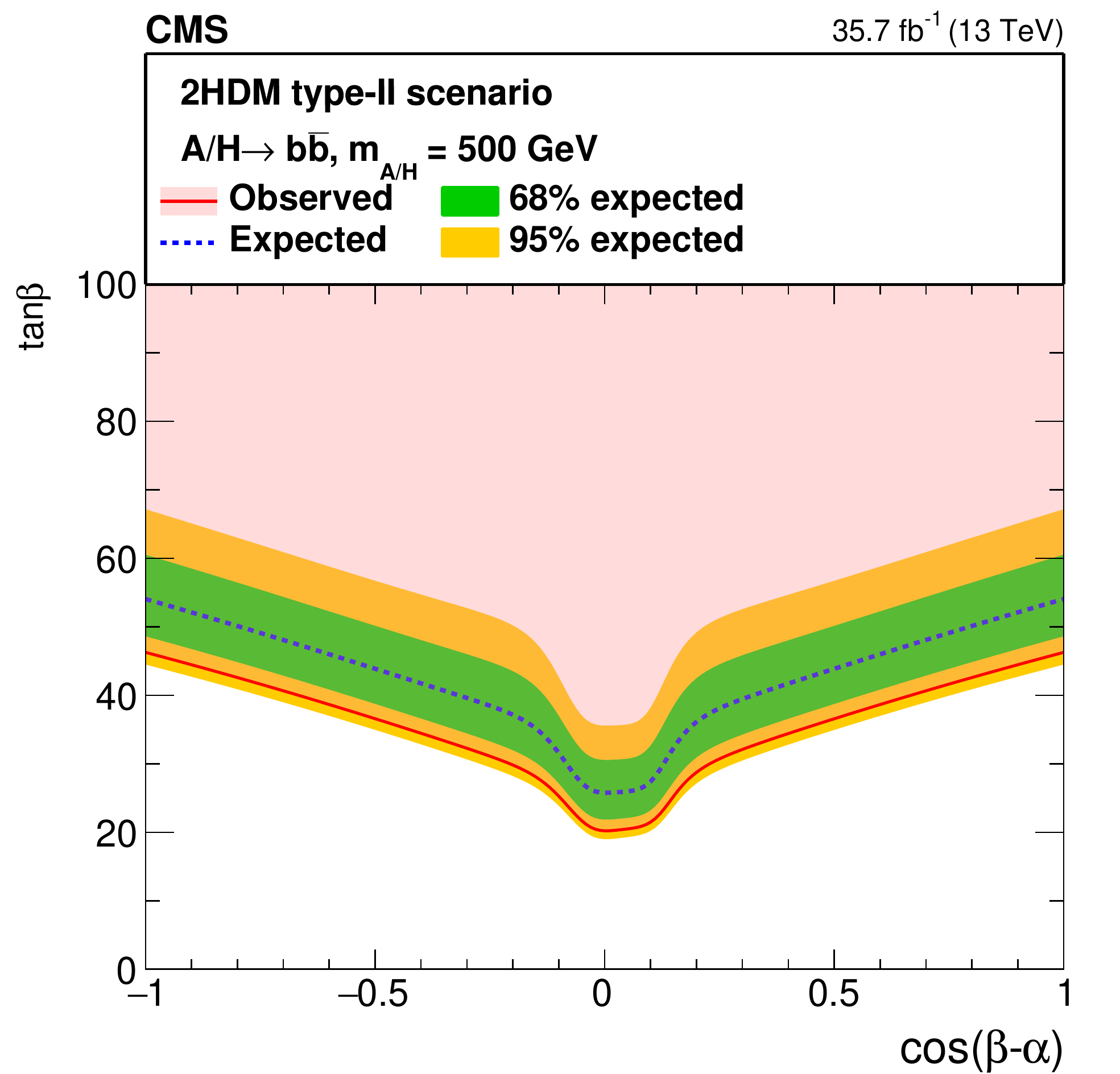}
	\caption{Upper limits for the parameter \tanb at 95\%
		confidence level for the flipped (left) and type-II
		(right) models as a function of \cosba
		in the full range of $[-1.0,1.0]$, for the mass $m_\PH=m_\PSA=500\GeV$.
		The inner and outer bands indicate the regions containing 68 and 95\%, respectively, of the distribution of limits expected under the background-only hypothesis.
	}
	\label{Fig:Interpretation:ULBrazil2HDM_500}
\end{figure}

The results for the type-II and flipped models are displayed
in Fig.~\ref{Fig:Interpretation:ULBrazil2HDM} as upper limits for
\tanb as a function of \cosba. Observed upper limits derived
from the ATLAS $\PSA \to \PZ\Ph$ analysis~\cite{Aaboud:2017sjh}
at a centre-of-mass energy of 13\TeV are shown as well.
The results for the flipped model presented here provide competitive
upper limits in the central region of \cosba and strong unique constraints on
\tanb.
Figure~\ref{Fig:Interpretation:ULBrazil2HDM_500} shows the upper limits for \tanb
as a function of \cosba in the type-II and flipped models for $\mAH = 500\GeV$.

\section{Summary}
A search for a heavy Higgs boson decaying into a bottom quark-antiquark pair and
accompanied by at least one additional bottom quark has been performed. The data analyzed
correspond to an integrated luminosity of 35.7\fbinv, recorded
in proton-proton collisions at a centre-of-mass energy
of $\sqrt{s} = 13\TeV$ at the LHC.
For this purpose, dedicated triggers using all-hadronic jet signatures combined with online \cPqb\ tagging were developed.
The signal is characterized by events with at least three b-tagged jets. The search has been performed in the invariant mass spectrum of the two leading jets that are also required to be b-tagged.

No evidence for a signal is found.
Upper limits on the Higgs boson cross
section times branching fraction are obtained in the mass region
300--1300\GeV at 95\% confidence level. They range from about 20\unit{pb} at the lower end of the mass range, to about 0.4\unit{pb} at 1100\GeV, and extend to considerably higher masses than those accessible to previous analyses in this channel.

The results are interpreted within various benchmark
scenarios of the minimal supersymmetric extension of the standard model (MSSM). They yield upper limits on the model parameter \tanb
as a function of the mass parameter \mA.
The observed limit at 95\% confidence level for \tanb is as low as about 25
at the lowest \mA value of 300\GeV in the $\mhmodp$ scenario with a higgsino mass parameter of $\mu=+200\GeV$.
In the hMSSM, scenarios with \tanb\ values above 22 to 60 for Higgs boson masses from 300 to 900\GeV are excluded at 95\% confidence level.
The results are also interpreted in the two Higgs doublet model (2HDM) type-II and flipped scenarios.
In the flipped 2HDM scenario, similar upper limits on \tanb as for the hMSSM are set over the full \cosba range and for Higgs boson masses from 300 to 850\GeV.
The limits obtained for the flipped scenario provide competitive
upper limits in the region around zero of $\cos(\beta-\alpha)$ and provide strong unique constraints on \tanb.

\begin{acknowledgments}
	
	The authors would like to thank Stefan Liebler and Oscar St\aa{}l for their help with the interpretation of the results in the 2HDM models.
	We congratulate our colleagues in the CERN accelerator departments for the excellent performance of the LHC and thank the technical and administrative staffs at CERN and at other CMS institutes for their contributions to the success of the CMS effort. In addition, we gratefully acknowledge the computing centres and personnel of the Worldwide LHC Computing Grid for delivering so effectively the computing infrastructure essential to our analyses. Finally, we acknowledge the enduring support for the construction and operation of the LHC and the CMS detector provided by the following funding agencies: BMWFW and FWF (Austria); FNRS and FWO (Belgium); CNPq, CAPES, FAPERJ, and FAPESP (Brazil); MES (Bulgaria); CERN; CAS, MoST, and NSFC (China); COLCIENCIAS (Colombia); MSES and CSF (Croatia); RPF (Cyprus); SENESCYT (Ecuador); MoER, ERC IUT, and ERDF (Estonia); Academy of Finland, MEC, and HIP (Finland); CEA and CNRS/IN2P3 (France); BMBF, DFG, and HGF (Germany); GSRT (Greece); NKFIA (Hungary); DAE and DST (India); IPM (Iran); SFI (Ireland); INFN (Italy); MSIP and NRF (Republic of Korea); LAS (Lithuania); MOE and UM (Malaysia); BUAP, CINVESTAV, CONACYT, LNS, SEP, and UASLP-FAI (Mexico); MBIE (New Zealand); PAEC (Pakistan); MSHE and NSC (Poland); FCT (Portugal); JINR (Dubna); MON, RosAtom, RAS, RFBR and RAEP (Russia); MESTD (Serbia); SEIDI, CPAN, PCTI and FEDER (Spain); Swiss Funding Agencies (Switzerland); MST (Taipei); ThEPCenter, IPST, STAR, and NSTDA (Thailand); TUBITAK and TAEK (Turkey); NASU and SFFR (Ukraine); STFC (United Kingdom); DOE and NSF (USA).
	
	\hyphenation{Rachada-pisek} Individuals have received support from the Marie-Curie programme and the European Research Council and Horizon 2020 Grant, contract No. 675440 (European Union); the Leventis Foundation; the A. P. Sloan Foundation; the Alexander von Humboldt Foundation; the Belgian Federal Science Policy Office; the Fonds pour la Formation \`a la Recherche dans l'Industrie et dans l'Agriculture (FRIA-Belgium); the Agentschap voor Innovatie door Wetenschap en Technologie (IWT-Belgium); the F.R.S.-FNRS and FWO (Belgium) under the ``Excellence of Science - EOS" - be.h project n. 30820817; the Ministry of Education, Youth and Sports (MEYS) of the Czech Republic; the Lend\"ulet ("Momentum") Programme and the J\'anos Bolyai Research Scholarship of the Hungarian Academy of Sciences, the New National Excellence Program \'UNKP, the NKFIA research grants 123842, 123959, 124845, 124850 and 125105 (Hungary); the Council of Science and Industrial Research, India; the HOMING PLUS programme of the Foundation for Polish Science, cofinanced from European Union, Regional Development Fund, the Mobility Plus programme of the Ministry of Science and Higher Education, the National Science Center (Poland), contracts Harmonia 2014/14/M/ST2/00428, Opus 2014/13/B/ST2/02543, 2014/15/B/ST2/03998, and 2015/19/B/ST2/02861, Sonata-bis 2012/07/E/ST2/01406; the National Priorities Research Program by Qatar National Research Fund; the Programa Estatal de Fomento de la Investigaci{\'o}n Cient{\'i}fica y T{\'e}cnica de Excelencia Mar\'{\i}a de Maeztu, grant MDM-2015-0509 and the Programa Severo Ochoa del Principado de Asturias; the Thalis and Aristeia programmes cofinanced by EU-ESF and the Greek NSRF; the Rachadapisek Sompot Fund for Postdoctoral Fellowship, Chulalongkorn University and the Chulalongkorn Academic into Its 2nd Century Project Advancement Project (Thailand); the Welch Foundation, contract C-1845; and the Weston Havens Foundation (USA).
	
\end{acknowledgments}

\bibliography{auto_generated}
\clearpage
\appendix

\section{Definition of Bukin function}
\label{sec:app:functions}

The Bukin function as implemented in ROOT version 6.06/01~\cite{Brun:1997pa} is defined as:
\begin{flalign}
	f(\mjj) &= A_p \exp \left[ -\ln 2 \frac{\ln^2\left(1+\sqrt{2}\xi \sqrt{\xi^2 + 1} \frac{(\mjj - x_p)}{\sqrt{\ln 2}\sigma_p}\right)}{\ln^2\left(1+2\xi(\xi - \sqrt{\xi^2 + 1})  \right)}   \right],\notag\\
	&\text{ if $x_1 < \mjj < x_2$,}\\
	f(\mjj) &= A_p \exp \left[ \pm \frac{\xi \sqrt{\xi^2 + 1} (\mjj - x_i) \sqrt{2 \ln 2}}{\sigma_p \ln(\sqrt{\xi^2 + 1} + \xi) \left( \sqrt{\xi^2 + 1} \mp \xi \right)^2 } + \rho_i \left( \frac{\mjj - x_i}{x_p - x_i}\right)^2 - \ln 2  \right],\notag\\
	&\text{ if $\mjj \leq x_1$ or $\mjj \geq x_2$},
\end{flalign}
where $\rho_i = \rho_1$ and $x_i = x_1$ for $\mjj \leq x_1$, $\rho_i = \rho_2$ and $x_i = x_2$ when $\mjj \geq x_2$, and:
\begin{equation}
x_{1,2} = x_p + \sigma_p \sqrt{2 \ln 2} \left( \frac{\xi}{\sqrt{\xi +1}} \mp 1 \right).
\end{equation}
The parameters $x_p$ and $\sigma_p$ are the peak position and width, respectively, and $\xi$ is an asymmetry parameter.

\section{Exclusion limits} \label{sec:app:limits}

The model-independent 95\% \CL\ limits on $\sigma(\Pp\Pp\to\PQb \PSA/\PH+\mathrm{X})\,\mathcal{B}(\PSA/\PH\to\bbbar)$ are listed in
Table~\ref{tab:limits:xsec} for different Higgs boson masses \mAH.
The 95\% \CL\ limits of $(\tanb,\mA)$ are listed in Tables~\ref{tab:limits:mhmodp}
to~\ref{tab:limits:lightstop} for different MSSM benchmark scenarios.

\begin{table}[htb]
	\centering
	\topcaption{Expected and observed 95\% \CL\  upper limits on $\sigma(\Pp\Pp\to\PQb \PSA/\PH+\mathrm{X})\,\mathcal{B}(\PSA/\PH\to\bbbar)$ in pb as a function of \mAH.}
	\begin{tabular}{crrrrrr}
		\hline
		Mass [GeV] &   $-2\sigma$ &   $-1\sigma$ & Median &   $+1\sigma$ &   $+2\sigma$ & Observed \\
		\hline
		300      &   10.8 &   14.3 &    19.7 &  27.5 &  36.5 &     19.1 \\
		350      &   6.3 &   8.4 &    11.7 &   16.3 &   21.7 &      14.0 \\
		400      &   3.6 &   4.8 &    6.7 &   9.2 &   12.3 &      5.7 \\
		500 &    1.7 &    2.2 &    3.1 &    4.4 &    5.9 &    1.9 \\
		600 &    1.0 &    1.4 &    1.9 &    2.7 &    3.7 &    2.1 \\
		700 &    0.7 &    0.9 &    1.3 &    1.8 &    2.4 &    1.5 \\
		900 &    0.4 &    0.6 &    0.8 &    1.2 &    1.6 &    0.9 \\
		1100&   0.36 &   0.49 &   0.68 &   0.96 &   1.36 &   0.40 \\
		1300&   0.36 &   0.48 &   0.68 &   0.96 &   1.31 &   0.50\\
		\hline
	\end{tabular}
	\label{tab:limits:xsec}
\end{table}

\begin{table}[htb]
	\centering
	\topcaption{Expected and observed 95\% \CL\  upper limits on \tanb as a function of \mA in the \mhmodp,
		$\mu = +200\GeV$, benchmark scenario. Since theoretical predictions
		for $\tanb > 60$ are not reliable, entries for which \tanb would exceed this value are indicated by \NA.}
	\begin{tabular}{crrcrrr}
		\hline
		Mass [GeV] &  $-2\sigma$ &  $-1\sigma$ & Median &  $+1\sigma$ &  $+2\sigma$ & Observed \\
		\hline
		300 &   19.3 &   22.0 &   25.8 &   30.6 &   35.7 &   25.4 \\
		350 &   21.5 &   24.4 &   28.5 &   33.6 &   39.0 &   31.2 \\
		400 &   22.6 &   25.5 &   29.4 &   34.4 &   39.7 &   27.3 \\
		500 &   26.9 &   30.2 &   34.9 &   40.9 &   47.4 &   28.3 \\
		600 &   32.9 &   37.1 &   43.0 &   50.6 &   58.5 &   44.5 \\
		700 &   39.0 &   44.2 &   51.7 &   \NA &   \NA &   55.9 \\
		900 &   58.5 &   \NA &   \NA &   \NA &  \NA &   \NA \\
		\hline
	\end{tabular}
	\label{tab:limits:mhmodp}
\end{table}

\begin{table}[htb]
	\centering
	\topcaption{Expected and observed 95\% \CL\  upper limits on \tanb as a function of \mA in the hMSSM
		benchmark scenario. Since theoretical predictions
		for $\tanb > 60$ are not reliable, entries for which \tanb would exceed this value are indicated by \NA.}
	\begin{tabular}{crrcrrr}
		\hline
		Mass [GeV] &  $-2\sigma$ &  $-1\sigma$ & Median &  $+1\sigma$ &  $+2\sigma$ & Observed \\
		\hline
		300 &   16.8 &   19.3 &   22.6 &   26.7 &   30.9 &   22.3 \\
		350 &   17.5 &   20.2 &   23.8 &   28.2 &   32.5 &   26.1 \\
		400 &   17.6 &   20.3 &   23.8 &   28.1 &   32.4 &   21.9 \\
		500 &   19.6 &   22.6 &   26.7 &   31.6 &   36.9 &   20.9 \\
		600 &   23.6 &   27.2 &   32.1 &   38.0 &   44.3 &   33.2 \\
		700 &   27.9 &   32.2 &   38.0 &   45.1 &   52.4 &   41.2 \\
		900 &   42.8 &   49.4 &   58.4 &   \NA &   \NA &   \NA \\
		\hline
	\end{tabular}
	\label{tab:limits:hmssm}
\end{table}

\begin{table}[htb]
	\centering
	\topcaption{Expected and observed 95\% \CL\ upper limits on \tanb as a function of \mA in the light $\sTau$ benchmark scenario. Since theoretical predictions
		for $\tanb > 60$ are not reliable, entries for which \tanb would exceed this value are indicated by \NA.}
	\begin{tabular}{crrcrrr}
		\hline
		Mass [GeV] &  $-2\sigma$ &  $-1\sigma$ & Median &  $+1\sigma$ &  $+2\sigma$ & Observed \\
		\hline
		300 &   19.9 &   23.6 &   28.8 &   35.8 &   43.7 &   28.2 \\
		350 &   21.0 &   25.0 &   30.8 &   38.4 &   47.5 &   34.7 \\
		400 &   21.7 &   25.5 &   31.2 &   38.8 &   47.9 &   28.0 \\
		500 &   25.0 &   29.8 &   37.2 &   47.8 &   \NA &   27.0 \\
		600 &   31.5 &   38.0 &   48.5 &   \NA &   \NA &   51.5 \\
		700 &   40.0 &   48.8 &   \NA &   \NA &   \NA &   \NA \\
		\hline
	\end{tabular}
	\label{tab:limits:lighstau}
\end{table}

\begin{table}[htb]
	\centering
	\topcaption{Expected and observed 95\% \CL\ upper limits on \tanb as a function of \mA in the light $\sTop$ benchmark scenario. Since theoretical predictions
		for $\tanb > 60$ are not reliable, entries for which \tanb would exceed this value are indicated by \NA.}
	\begin{tabular}{crrcrrr}
		\hline
		Mass [GeV] &  $-2\sigma$ &  $-1\sigma$ & Median &  $+1\sigma$ &  $+2\sigma$ & Observed \\
		\hline
		300 &   22.2 &   26.9 &   34.6 &   46.3 &   \NA &   33.6 \\
		350 &   23.6 &   28.9 &   37.6 &   52.3 &   \NA &   44.5 \\
		400 &   23.8 &   29.3 &   37.9 &   51.9 &   \NA &   32.9 \\
		500 &   27.9 &   34.8 &   47.4 &   \NA &   \NA &   30.7 \\
		600 &   37.4 &   49.0 &   \NA &   \NA &   \NA &   \NA \\
		\hline
	\end{tabular}
	\label{tab:limits:lightstop}
\end{table}

\cleardoublepage \section{The CMS Collaboration \label{app:collab}}\begin{sloppypar}\hyphenpenalty=5000\widowpenalty=500\clubpenalty=5000\vskip\cmsinstskip
\textbf{Yerevan Physics Institute, Yerevan, Armenia}\\*[0pt]
A.M.~Sirunyan, A.~Tumasyan
\vskip\cmsinstskip
\textbf{Institut f\"{u}r Hochenergiephysik, Wien, Austria}\\*[0pt]
W.~Adam, F.~Ambrogi, E.~Asilar, T.~Bergauer, J.~Brandstetter, E.~Brondolin, M.~Dragicevic, J.~Er\"{o}, A.~Escalante~Del~Valle, M.~Flechl, R.~Fr\"{u}hwirth\cmsAuthorMark{1}, V.M.~Ghete, J.~Hrubec, M.~Jeitler\cmsAuthorMark{1}, N.~Krammer, I.~Kr\"{a}tschmer, D.~Liko, T.~Madlener, I.~Mikulec, N.~Rad, H.~Rohringer, J.~Schieck\cmsAuthorMark{1}, R.~Sch\"{o}fbeck, M.~Spanring, D.~Spitzbart, A.~Taurok, W.~Waltenberger, J.~Wittmann, C.-E.~Wulz\cmsAuthorMark{1}, M.~Zarucki
% ...
\vskip\cmsinstskip
% ...
1:  Also at Vienna University of Technology, Vienna, Austria\\
% ...
\end{sloppypar}
\end{document}